\documentclass[acmsmall,screen]{acmart}

\usepackage{multirow,makecell}
\usepackage{tcolorbox}
\usepackage{color}
\usepackage{xcolor}
\usepackage{colortbl}
\usepackage{listings,amsfonts}
\usepackage{caption}
\usepackage{subcaption}
\usepackage{threeparttable}
\usepackage{bbding}
\usepackage{graphicx}
\usepackage{booktabs} 
\usepackage{enumitem}
\usepackage{longtable}
\usepackage{seqsplit}
\usepackage{pifont}
\usepackage[linesnumbered,ruled,vlined]{algorithm2e}
\usepackage{setspace}
\usepackage{wrapfig}

\definecolor{customblue}{HTML}{006ca6}
\definecolor{customgreen}{HTML}{009264}
\definecolor{custombrown}{HTML}{ff3d00}
\AtEndPreamble{
 \usepackage{hyperref}

 \hypersetup{
  colorlinks = true,
  linkcolor = customblue,
  anchorcolor = customblue,
  citecolor = customblue,
  filecolor = customblue,
  urlcolor = customblue
 }
}

\newcommand{\tool}[1]{\textsc{LogicHunter}}

\title{\tool{}: Testing LLM Agent Frameworks with an Agentic Oracle}

\begin{document}


\author[M. Long]{Minghui Long}
\orcid{0009-0000-4196-2474}
\email{mhlong@hust.edu.cn}
\authornotemark[1]
\affiliation{%
  \department{Hubei Key Laboratory of Distributed System Security}
  \department{Hubei Engineering Research Center on Big Data Security}
  \department{School of Cyber Science and Engineering}
  \institution{Huazhong University of Science and Technology}
  \city{Wuhan}           
  \country{China}
}

\author[Y. Zhao]{Yanjie Zhao}
\orcid{0000-0001-8793-5367}
\email{Yanjie_Zhao@hust.edu.cn}
\authornote{Minghui Long and Yanjie Zhao are the co-first authors.}
\affiliation{%
  \department{Hubei Key Laboratory of Distributed System Security}
  \department{Hubei Engineering Research Center on Big Data Security}
  \department{School of Cyber Science and Engineering}
  \institution{Huazhong University of Science and Technology}
  \city{Wuhan}           
  \country{China}
}

\author[H. Wang]{Haoyu Wang}
\orcid{0000-0003-1100-8633}
\email{haoyuwang@hust.edu.cn}
\authornote{Haoyu Wang (haoyuwang@hust.edu.cn) is the corresponding author.}
\affiliation{%
  \department{Hubei Key Laboratory of Distributed System Security}
  \department{Hubei Engineering Research Center on Big Data Security}
  \department{School of Cyber Science and Engineering}
  \institution{Huazhong University of Science and Technology}
  \city{Wuhan}           
  \country{China}
}

\begin{abstract}

Large Language Model (LLM) agent frameworks such as LangChain, LlamaIndex, and CrewAI have become critical infrastructure powering production AI systems, yet they remain severely under-tested due to fundamental challenges in automated testing. Unlike traditional software, where crashes serve as reliable oracles, defects in these pure Python frameworks manifest as ordinary exceptions or silent semantic failures, creating profound oracle ambiguity. This problem is exacerbated by strict type governance through Pydantic schemas and complex protocol requirements that cause existing fuzzers to generate overwhelming invalid inputs, while traditional test generators produce only trivial cases with weak regression assertions.

We present \tool{}, a fuzzing framework that addresses both the generation and oracle challenges through active specification-aware testing. \tool{} employs specification-driven generation that systematically fuses formal type constraints with authentic usage patterns from real-world repositories, synthesizing inputs that are valid by construction yet semantically extreme, equipped with behavioral probes to expose silent failures. To resolve oracle ambiguity, we introduce the Agentic Oracle, which transcends passive classification by actively retrieving documentation, navigating source code, and inspecting runtime states through a ReAct-based architecture with Dual-Layer State Management and Dual-Stream Memory. Evaluated on three widely deployed frameworks, \tool{} discovered 40 previously unknown bugs with 30 confirmed and 26 fixed by developers, while state-of-the-art baselines reported no bugs as final findings. The Agentic Oracle achieves  91.17\% precision, surpassing the best passive approach at 29.27\% by 61 percentage points.

\end{abstract}

\maketitle

\section{Introduction}
\label{sec:introduction}

Large language model (LLM) \textbf{agent frameworks} such as LangChain~\cite{Chase_LangChain_2022}, LlamaIndex~\cite{Liu_LlamaIndex_2022}, and CrewAI~\cite{crewai2025} have emerged as the foundational infrastructure for building modern AI systems~\cite{xi2023agent,Wang2024agent,luo2025agent}. Their popularity is reflected not only in widespread industrial adoption but also in unprecedented development activity: LangChain alone has accumulated over 30,000 GitHub issues~\cite{Chase_LangChain_2022}, signaling both rapid evolution and persistent reliability challenges. Despite their critical role in production pipelines, agent frameworks remain largely under-tested, and systematic techniques for discovering bugs in this domain are still lacking.

\textbf{The core challenge lies in the test oracle problem.} Traditional automated testing techniques rely on clear, cheap oracles: crashes for native-code fuzzers, differential outputs for numerical libraries, or metamorphic relations for transformation-based systems~\cite{oracle2014,Metamorphic_survey}. In the studied agent frameworks, which are implemented almost entirely in pure Python, defects manifest as ordinary exceptions (e.g., \texttt{ValueError}, \texttt{KeyError}) or, more insidiously, as silent semantic errors where execution completes successfully but violates intended behavior. The same exception type may arise from legitimate API misuse or from internal logic errors, making exception-based oracles fundamentally ambiguous. Silent failures further complicate matters, as incorrect behavior may raise no exception at all.

\textbf{This ambiguity renders existing testing tools ineffective.} Search-based unit test generators~\cite{pynguin,codamosa,telpa} target code coverage with weak regression assertions, generating tests that pass rather than exposing semantic bugs. Traditional fuzzers~\cite{atheris2020,li2023pyrtfuzz} emphasize input diversity through mutation but produce overwhelming noise: invalid inputs that violate schema constraints enforced by Pydantic models~\cite{Colvin_Pydantic_2025} or protocol requirements, triggering ordinary exceptions indistinguishable from genuine bugs. Recent work has explored LLMs as test oracles~\cite{milev2025toolfuzzautomatedagent,oracleinllm,2023judge}, but faces critical limitations when applied to rapidly evolving Python frameworks: knowledge obsolescence as APIs change, opaque reasoning that undermines trust~\cite{huang2024largelanguagemodelsselfcorrect}, and passive operation on fixed snapshots that cannot actively inspect runtime state.

\textbf{Beyond oracles, the generation challenge is equally fundamental.} Agent frameworks expose complex APIs with type constraints and protocol requirements. Blindly mutating inputs (as fuzzers do) yields overwhelming invalid cases (e.g., violating Pydantic schemas, type mismatches, or missing required fields) that trigger ordinary exceptions indistinguishable from genuine bugs. Conversely, simply satisfying type constraints is insufficient: valid yet trivial inputs (e.g., empty strings, default values) rarely expose logic errors. What is needed are inputs that are strictly specification-compliant yet semantically extreme, systematically exploring boundary conditions within the usage space, which is a capability current generators lack.

\textbf{This work addresses the generation-oracle gap through a unified framework.} We propose \tool{}, which leverages formal specifications embedded in agent frameworks (e.g., type hints, Pydantic schemas) to synthesize inputs that are valid-by-construction yet designed to stress corner cases. We introduce an \textbf{Agentic Oracle} that actively retrieves documentation, inspects runtime states, and performs targeted executions to gather evidence before rendering verdicts. This design enables principled diagnosis of both exception-triggering and silent semantic bugs while maintaining cost-effectiveness through closed-loop architecture.

In summary, our contributions are:

\begin{itemize}[leftmargin=1em]
\item We identify the oracle problem as a central obstacle in testing LLM agent framework implementations and analyze the limitations of existing techniques.

\item We present \tool{}, a fuzzing framework that unifies specification-driven generation with an Agentic Oracle for active diagnosis through documentation retrieval, runtime inspection, and targeted execution.

\item We evaluate \tool{} on three widely adopted agent frameworks (LangChain, LlamaIndex, CrewAI), discovering 40 previously unknown bugs with 30 developer confirmations and 26 fixes. The Agentic Oracle achieves 91.17\% precision, outperforming state-of-the-art baselines by 61 percentage points, while baselines discovered zero bugs.
\end{itemize}

\section{Background and Motivation}
\label{sec:background}

\subsection{Agents in Software Engineering}
\label{ssec:agents_se}

LLM-based agents have been adopted across software engineering tasks including code generation, automated repair, bug localization, test generation, and issue resolution~\cite{agent4se,swe-agent,llm4se}. Unlike single-shot LLM invocations, agents operate through iterative thought-action-result loops, where intermediate reasoning guides tool invocation and subsequent decisions based on environmental feedback~\cite{yao2023react,toolformer}.
This paradigm enables sustained interaction with development environments such as code editors, compilers, debuggers, and test frameworks. Recent systems demonstrate that agents can autonomously navigate repositories, modify code, run tests, and iteratively refine solutions, achieving promising results on realistic maintenance tasks~\cite{swe-bench,metagpt}.

At the same time, empirical studies analyzing execution traces reveal that many agent failures do not manifest as crashes but as semantically incorrect yet well-formed behaviors~\cite{toolllm,intercode}, highlighting the need to test the underlying framework infrastructure on which such agent behaviors depend.

\subsection{Agent Frameworks as a Testing Context}
\label{ssec:framework_characteristics}

From a software testing perspective, LLM agent frameworks do not introduce a fundamentally new class of low-level defects. Recent empirical studies~\cite{xue_agentstudy} on LLM agent framework bugs support this observation: framework-level bugs can still be classified into conventional software-engineering root causes, including API misuse, incompatibility, assignment issues, parameter/argument issues, code logic issues, import errors, typos, exception-handling errors, and numerical computation errors. This is expected, since agent-specific logic such as reasoning-path selection, task planning, and tool-use decisions mainly emerges in applications built on top of these frameworks, while the frameworks themselves are implemented as conventional software infrastructure.

These conventional defects become particularly important in the agent-system context. Agent applications depend on framework components for tool parsing, memory, retrieval, callbacks, external-service integration, and workflow orchestration. Xue et al. report that bugs in LLM agent frameworks can propagate to downstream applications, causing incorrect task execution, external API integration failures, context-management breakdowns, downstream errors, and crashes ~\cite{xue_agentstudy}. Recent studies on multi-agent systems further show that failures are difficult to localize: failure attribution remains underexplored and labor-intensive, and existing methods achieve limited accuracy when identifying the responsible failure step ~\cite{zhang2025icml-agent}. This motivates testing techniques that expose ordinary-looking implementation defects before they propagate into downstream agent-system failures.

Understanding the following properties motivates our technical approach.

\subsubsection{Architectural Characteristics}

\textbf{Strict Type Governance.} Agent frameworks extensively leverage Pydantic~\cite{Colvin_Pydantic_2025} for runtime validation, enforcing complex nested schemas with inheritance hierarchies and custom validators. Unlike simple type hints, these constraints are actively checked at instantiation and method invocation, rejecting inputs that violate schema invariants before reaching core logic.

\textbf{Protocol-Oriented Design.} Frameworks define extensibility through abstract base classes (e.g., \texttt{BaseTool}, \texttt{BaseMemory}) requiring specific initialization sequences and state management protocols. Correct usage demands not just syntactically valid calls but proper object orchestration following implicit conventions.

\textbf{Pervasive Asynchrony.} To accommodate LLM latency, most APIs support both synchronous and asynchronous execution paths with complex event loop management. This duality multiplies the testing surface and introduces concurrency-related failure modes.

\subsubsection{Semantic Characteristics}

\textbf{Configuration-Driven Behavior.} Compared with simple functional libraries, agent frameworks expose highly parameterized APIs where behavior is determined by configuration objects, callback chains, and dynamic prompt templates. The same code path can exhibit vastly different semantics depending on runtime configuration.

\textbf{External Service Dependencies.} Agent frameworks frequently interact with LLM APIs, vector databases, and web services. While testing typically mocks these dependencies, the frameworks must handle diverse error modes from external systems (rate limits, timeout, malformed responses), expanding the space of valid but edge-case behaviors.

\textbf{Rapid API Evolution.} Active development introduces frequent breaking changes, deprecations, and behavioral modifications. Documentation lags behind implementation, and parametric knowledge in pre-trained models quickly becomes obsolete.


\subsection{Why Existing Techniques Fail}
\label{ssec:why_fail}

In this testing context, the above characteristics expose a mismatch between existing tools and framework-level defect discovery. Strict type governance and protocol requirements (\autoref{ssec:framework_characteristics}) demand specification-aware generation, yet existing tools explore inputs blindly. Configuration-driven behavior and external dependencies introduce subtle semantic bugs that manifest as silent failures, yet current oracles lack mechanisms to detect them. Rapid API evolution invalidates static knowledge, yet LLM-based approaches rely on parametric memory. 
The oracle problem is amplified in agent frameworks because neither crashes nor differential testing provide reliable signals. This fundamental challenge cascades through both test generation and verification. We now detail six challenges that emerge from this mismatch between framework properties and tool capabilities.

\subsubsection{Test Generation Challenges}

\textbf{Challenge 1 (C1): Specification Compliance.} Search-based tools~\cite{pynguin} and traditional fuzzers~\cite{atheris2020,li2023pyrtfuzz} explore inputs through random mutation and evolutionary search. However, \textbf{agent frameworks reject vast swathes of the input space through schema validation before reaching testable logic}. Naive mutation produces overwhelming false positives: syntactically valid Python that violates Pydantic constraints, protocol requirements, or type invariants.
While similar validation exists in frameworks like FastAPI~\cite{Ramirez_FastAPI}, agent frameworks compound this with protocol requirements: valid types must also follow implicit initialization sequences and state management conventions. Existing tools lack mechanisms to learn and respect these multi-layered constraints.

\textbf{Challenge 2 (C2): Scalability for Expensive Oracles.} LLM-based oracles are computationally expensive, processing only tens of test cases feasibly. However, traditional fuzzers generate thousands of inputs, most failing trivially or passing obviously. \textit{There is a fundamental mismatch}: expensive semantic analysis requires pre-filtered, high-value inputs, but existing generators lack the semantic awareness to produce them. Unit test generators prioritize coverage and regression, producing tests optimized for passing rather than bug exposure.

\begin{figure}[htbp]
\centering
\includegraphics[width=\linewidth]{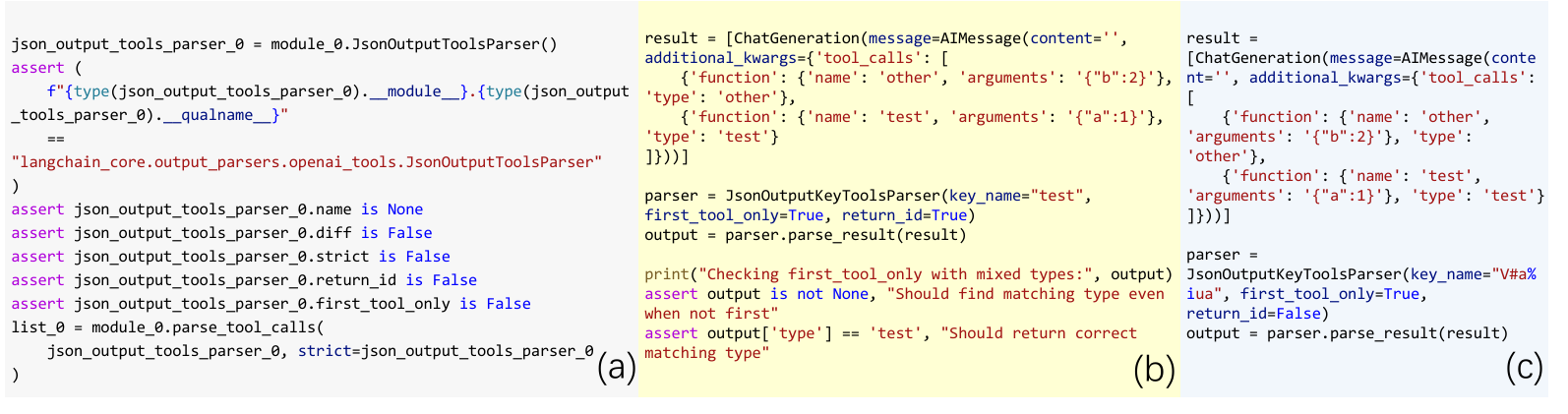}
\caption{\textbf{Comparison of test generation approaches.} 
(a) Unit-Test Generation produces valid inputs with regression-oriented assertions. 
(b) Probe-Oriented Testing (our approach) generates spec-compliant inputs with expectation-oriented behavioral probes. 
(c) Fuzzing explores semantically blind inputs without assertions.}
\label{fig:challenge_example_code}
\end{figure}

\textbf{Challenge 3 (C3): Silent Failure Detection.} Many logic bugs produce well-typed outputs that violate intended semantics. As illustrated in \autoref{fig:challenge_example_code}, fuzzers lack assertions entirely (c), while unit test generators produce regression checks (a) that cannot distinguish silent errors from correct behavior~\cite{konstantinou2024oracles}. Detecting silent failures requires behavioral probes (b): expectation-oriented assertions encoding intended behavior, enabling identification of suspicious executions from a large set of passing tests. Existing tools neither generate nor prioritize such probes.

\subsubsection{Test Oracle Challenges}

\textbf{Challenge 4 (C4): Knowledge Obsolescence and Opaque Reasoning.} LLM-based oracles relying on parametric knowledge face two critical risks. \textit{First, agent frameworks evolve rapidly}; static models judge based on outdated APIs, flagging valid new features as bugs. \textit{Second, the inference process remains a black box}: even correct verdicts lack verifiable rationale. Due to stochastic nature, verdicts are fragile: minor prompt variations or model updates flip decisions. Without external evidence, users cannot distinguish reasoned diagnosis from confident hallucination.

\textbf{Challenge 5 (C5): Context Saturation.} To mitigate hallucination, naive approaches inject all potentially relevant context (documentation, stack traces, source code) into prompts. This creates \textit{context saturation}: real-world error logs and documentation exceed effective context windows. Even within token limits, the ``Lost-in-the-Middle'' phenomenon~\cite{liu-etal-2024-lost} disperses attention, causing models to overlook critical failure details buried in verbose inputs. Indiscriminate context stuffing paradoxically degrades accuracy.

\textbf{Challenge 6 (C6): Passive Observation Bottleneck.} Current oracles operate as passive classifiers, restricted to pre-determined information snapshots. Diagnosing deep semantic bugs often hinges on hidden states within call chains (e.g., specific variable values in private methods). Since relevant runtime states cannot be determined a priori, passive methods face a dilemma: dump entire execution traces (triggering C5) or provide insufficient data. Inability to actively verify hypotheses prevents models from leveraging their full reasoning potential.

\subsection{Design Philosophy and Key Insights}
\label{ssec:approach_preview}

The six challenges stem from a mismatch between available API cues, semantic failure modes, and passive testing tools. We address this through two key insights.

\textbf{Insight 1: Specifications as Generation Oracles.} Agent frameworks declare valid inputs through Pydantic schemas, type hints, and docstrings, while real-world repositories demonstrate correct API orchestration. By mining and combining both sources, we bias generation toward executable, semantically challenging tests, reducing trivial validation failures before oracle analysis.

\textbf{Insight 2: Active Reasoning for Semantic Diagnosis.} Distinguishing genuine defects from API misuse requires inspecting implementation states and cross-referencing documentation. An oracle equipped with code navigation and execution tools can actively formulate hypotheses and gather targeted evidence, enabling dynamic documentation retrieval (C4), selective information gathering (C5), and on-demand state inspection (C6).

These insights lead to a closed-loop architecture where specification-driven generation produces high-value inputs while agentic verification performs evidence-based diagnosis. We operationalize this in \tool{}, described next.

\section{Approach}
\label{sec:approach}

We present \tool{}, a fuzzing framework that bridges the generation-oracle gap through specification-aware testing and tool-augmented reasoning.

\begin{figure}[htbp]
    \centering
    \includegraphics[width=0.9\linewidth]{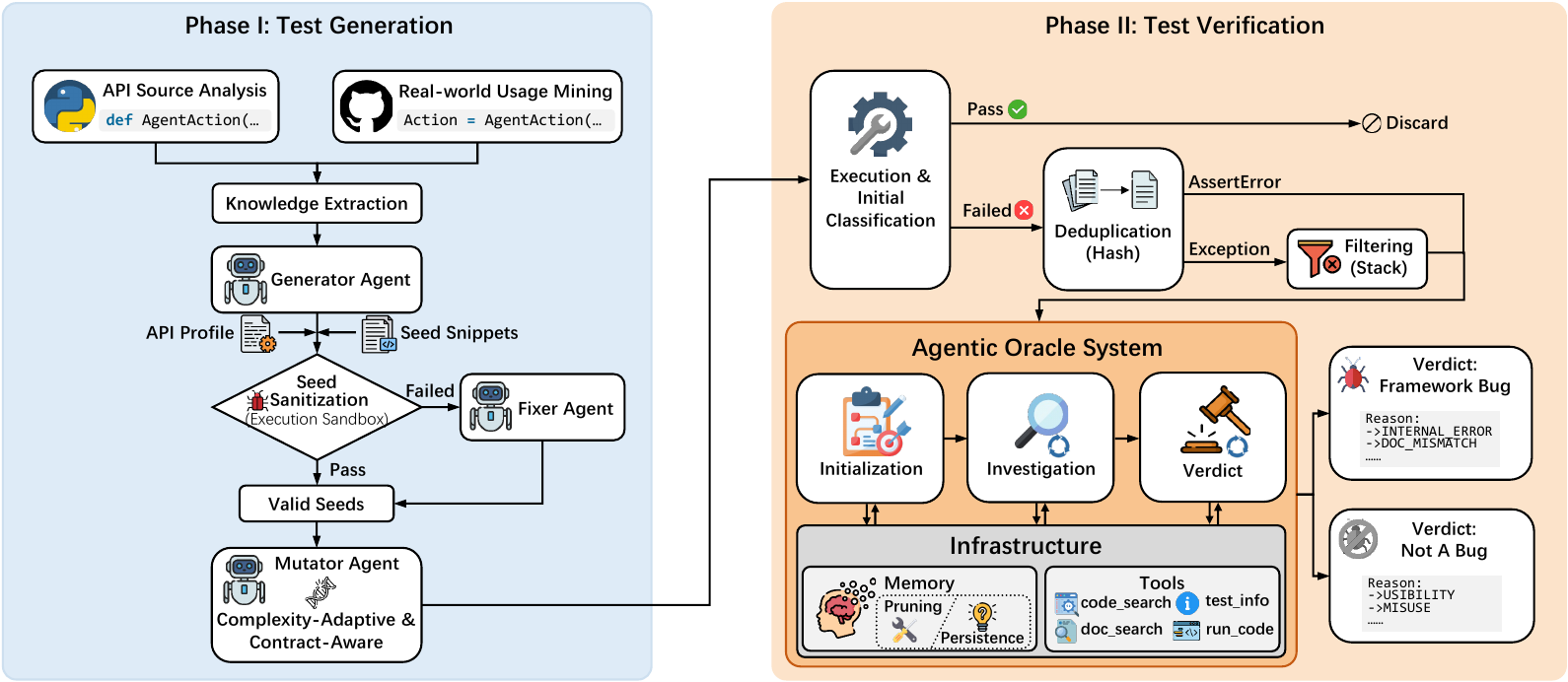}
    \caption{The workflow of \tool{}. Phase I synthesizes specification-compliant tests with behavioral probes. Phase II filters anomalies and deploys the Agentic Oracle for evidence-based diagnosis.}
    \label{fig:workflow}
\end{figure}

\subsection{System Overview}
\label{ssec:overview}

\tool{} implements a two-phase pipeline (see \autoref{fig:workflow}) that transforms the six challenges into technical solutions:

\textbf{Phase I: Test Generation} (\autoref{ssec:generation}) Test Generation (§ 3.2) combines source-level explicit contracts with usage-derived implicit protocols. A Generator Agent produces executable seeds and API Profiles; a Fix Agent repairs invalid seeds through execution; and a Mutator Agent generates diverse, probe-equipped tests for exposing silent failures (C1--C3).

\textbf{Phase II: Test Verification} (\autoref{ssec:verification}) applies hash-based deduplication and validity filtering to extract unique anomalies. The \textbf{Agentic Oracle System} determines defects through investigation, operating under the ReAct\cite{yao2023react} paradigm with explicit state management and dual-stream memory. It inspects source code, retrieves documentation, and executes reproduction scripts to formulate verdicts. Anomalies are classified into a six-dimensional defect taxonomy, requiring high-confidence consensus before flagging bugs (addressing C4-C6).

\subsection{Test Generation}
\label{ssec:generation}

We design the generator to maximize downstream oracle utility along three dimensions, corresponding to the generation challenges C1--C3:

\textbf{High Validity.} Generated tests should adhere to both \textit{explicit contracts} (documentation, type hints, and schemas) and \textit{implicit protocols} (context construction patterns) derived from real-world usage, thereby reducing trivial validation failures before oracle analysis.

\textbf{High Semantic Complexity.} Generated tests should achieve dual-space coverage over both internal logic semantics (diverse control-flow branches) and documentation semantics (systematic exploration of the documented input space, even when multiple documented behaviors map to identical internal paths).

\textbf{High-Level Probe Signals.} Generated tests should include expectation-oriented assertions that act as behavioral probes over both implementation behavior and documented semantics, making silent failures observable for later oracle validation.

\subsubsection{Knowledge Extraction and Context Modeling}
\label{sssec:knowledge_extraction}

Agent frameworks expose explicit contracts through type hints, Pydantic schemas, and docstrings, while real-world repositories demonstrate implicit protocols for API orchestration. Mining both sources helps generate inputs that follow practical usage patterns while exploring semantic boundaries.

We implement an automated Knowledge Extraction phase bridging formal definitions and practical application:

\textbf{API Source Analysis.} We traverse the target framework's package structure, using static analysis to extract complete source code, function signatures, type hints, and docstrings. This provides source-level cues for interface constraints (explicit contracts).

\textbf{Real-World Usage Corpus.} To capture implicit protocols for object orchestration and state management, we mine high-quality repositories such as official integrations and top-rated framework projects. An AST-based parser extracts relevant snippets and maps APIs to invocation contexts. Although these snippets are often not directly executable, they provide templates for constructing realistic API contexts.

\begin{figure}[htbp]
  \centering
  \includegraphics[width=0.8\linewidth]{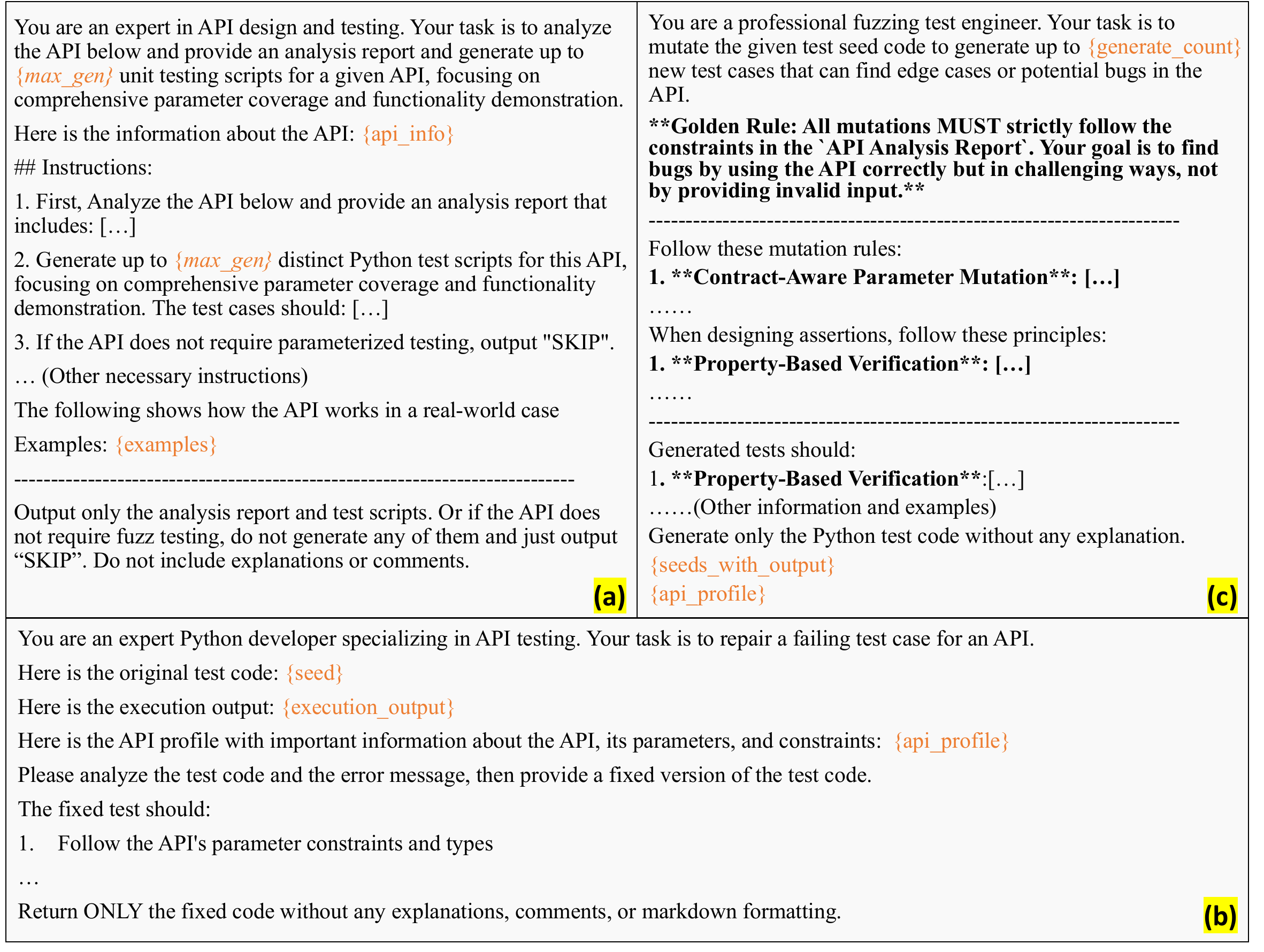}
  \caption{Prompt templates for (a) generator agent, (b) fix agent, and (c) mutator agent.}
  \label{fig:prompts}
\end{figure}

\subsubsection{Specification-Driven Seed Synthesis}
\label{sssec:seed_synthesis}

By fusing extracted specifications with usage templates through an LLM agent, we can simultaneously produce executable seeds and structured mutation blueprints, enabling efficient exploration of the semantic space while respecting constraints.

For each target API, the \textbf{Generator Agent} fuses extracted explicit contracts (source specifications) with mapped implicit protocols (usage templates) to produce foundational testing assets. Guided by a task-oriented few-shot prompt (\autoref{fig:prompts}(a)), the agent synthesizes two synchronized outputs:

\textbf{Seed Code Snippet.} A functional, executable script correctly instantiating the target API. This turns static templates into executable contexts by resolving dependencies and runtime setup.

\textbf{API Profile.} A structured mutation blueprint consisting of: (1) \textbf{API Complexity Score} $C(api)$, a quantitative metric for dynamic resource allocation; (2) \textbf{Documentation Context}, capturing input space semantics from docstrings and type hints; (3) \textbf{Logic Pseudocode}, a distilled representation of internal implementation. We use pseudocode rather than full source to enable logic reasoning while preventing context overflow and attention dispersion during generation.

\textbf{Seed Sanitization.} A \textbf{Fixer Agent} ensures mutation begins with a reliable baseline. It executes initial seeds and, upon failure, iteratively repairs them using common patterns (e.g., resolving ImportError, correcting Pydantic validation errors). Only seeds passing validation are promoted to Golden Seeds.

\subsubsection{Complexity-Aware Mutation and Behavioral probes Generation}
\label{sssec:mutation}

Rather than blindly mutating inputs, we guide mutation with API Profiles that combine documentation semantics and logic pseudocode, producing challenging edge cases together with behavioral probes for silent failures.

The \textbf{Mutator Agent} transforms Golden Seeds into a diverse test suite. Unlike bit-level fuzzing, this agent mutates tests at the API-usage level. Guided by a systematic prompt (\autoref{fig:prompts}(c)), it performs multidimensional mutations ranging from boundary parameter injection to control flow reordering while explicitly respecting type constraints and state dependencies. This strategy suppresses trivial invalid inputs, steering the LLM toward ``valid but challenging'' edge cases probing logical boundaries.

\textbf{Dynamic Resource Allocation.} We allocate mutation budgets by API complexity:
\[
N_{mut}=\lfloor \alpha \cdot C(api)/M_{max}\rfloor\cdot M_{max},
\]
where \(6 \le C(api) \le 30\), \(\alpha\) is a global scale factor, and \(M_{max}\) is the maximum mutations per LLM call. Thus, APIs with richer parameters, stronger constraints, or more complex logic receive larger budgets.

\textbf{Probe-Oriented Assertion Synthesis.} A key design choice is explicitly decoupling assertion
generation from the role of definitive oracle. We define a behavioral probe as an executable
assertion $p$ attached to a generated test $t$ that checks a local semantic property inferred from
the API profile, such as return-type consistency, field preservation, idempotency, or boundary
behavior. Formally, given an execution observation $o(t)$, the probe produces
$p(o(t)) \in \{\mathsf{pass}, \mathsf{anomaly}\}$. A probe-triggered anomaly is only a suspicious
signal and must be validated by the Agentic Oracle before being reported as a bug.

\subsection{Test Verification}
\label{ssec:verification}

The Test Verification phase bridges massive test generation and precise bug diagnosis. Since the Mutator Agent prioritizes semantic complexity over correctness, the raw test suite inevitably contains noise: redundancy (duplicate failures) and invalidity (hallucinated API usages). We propose a two-stage verification pipeline. The first stage (\autoref{sssec:preprocessing}) employs a deterministic algorithm sifting raw inputs into a compact set of unique, valid candidates. The second stage (\autoref{ssec:oracle}) deploys the Agentic Oracle to diagnose deep logic bugs.

\subsubsection{Execution, Deduplication, and Filtering}
\label{sssec:preprocessing}

Before invoking the expensive Agentic Oracle, we can efficiently eliminate redundant and invalid test cases through deterministic analysis of execution traces, focusing oracle attention on unique, high-potential anomalies.

This module rigorously cleanses the raw test suite, minimizing computational cost for the downstream oracle. We formalize this as a pipeline of three operations: Execution, Hash-based Deduplication, and Validity Filtering.

\begin{algorithm}[t]
\scriptsize  
\caption{Deterministic Verification Pipeline}
\label{alg:preprocessing}
\SetKwInOut{KwIn}{Input}
\SetKwInOut{KwOut}{Output}
\SetKwFunction{FExec}{Execute}
\SetKwFunction{FHash}{Hash}
\setstretch{0.9} 

\KwIn{Raw test suite $\mathcal{T}_{gen}$, Library Paths $\mathcal{L}$}
\KwOut{Unique valid cases $\mathcal{T}_{unique}$}

$\mathcal{T}_{fail} \gets \emptyset$; \quad $Map_{dedup} \gets \emptyset$; \quad $\mathcal{T}_{unique} \gets \emptyset$\;

\tcp{Step 1: Execution}
\ForEach{$t \in \mathcal{T}_{gen}$}{
    $res, trace \gets \FExec(t)$\;
    \lIf{$res \neq \text{PASS}$}{$\mathcal{T}_{fail} \gets \mathcal{T}_{fail} \cup \{(t, trace)\}$}
}

\tcp{Step 2: Deduplication}
\ForEach{$(t, trace) \in \mathcal{T}_{fail}$}{
    \uIf{$t \in \mathcal{T}_{exc}$}{
        $h \gets \FHash(trace.\text{TopFrame}().\text{Signature}())$\;
    }
    \uElseIf{$t \in \mathcal{T}_{asrt}$}{
        $h \gets \FHash(t.\text{GetTriggerStatement}().\text{Normalized}())$\;
    }
    \lIf{$h \notin Map_{dedup}$}{$Map_{dedup}[h] \gets (t, trace)$}
}

\tcp{Step 3: Validity Filtering}
\ForEach{$(t, trace) \in \text{Values}(Map_{dedup})$}{
    $isValid \gets \textbf{true}$\;
    \If{$t \in \mathcal{T}_{exc}$}{
        $f_0 \gets trace.\text{TopFrame}()$\;
        $hasLib \gets \exists f \in trace : f.\text{Path} \in \mathcal{L}$\;
        \lIf{$(f_0 \in t.\text{Path}) \lor (\neg hasLib)$}{$isValid \gets \textbf{false}$}
    }
    \lIf{$isValid$}{$\mathcal{T}_{unique} \gets \mathcal{T}_{unique} \cup \{t\}$}
}
\KwRet $\mathcal{T}_{unique}$\;
\end{algorithm}

\textbf{(1) Execution and Classification.} Let $\mathcal{T}_{gen}$ be the set of generated test cases. We execute each $t \in \mathcal{T}_{gen}$ in an isolated sandbox. Based on exit status, we partition $\mathcal{T}_{gen}$ into passed cases $\mathcal{T}_{pass}$ and failed cases $\mathcal{T}_{fail}$. Since $\mathcal{T}_{pass}$ implies no observable anomaly, we discard it and focus on $\mathcal{T}_{fail}$, further classified into:
\begin{itemize}[leftmargin=1em]
    \item \textbf{Runtime Exceptions ($\mathcal{T}_{exc}$):} Cases terminated by unhandled exceptions (e.g., \texttt{TypeError}), indicating potential crashes or invalid usages.
    \item \textbf{Assertion Failures ($\mathcal{T}_{asrt}$):} Cases where execution violated a generated semantic assertion, indicating potential silent logic errors.
\end{itemize}

\textbf{(2) Hash-Based Deduplication.} \textbf{To eliminate redundant failures caused by identical root causes, we apply deduplication before filtering}, ensuring we do not waste resources analyzing validity of duplicate traces. We define a hash mapping $\Phi: \mathcal{T}_{fail} \to \mathbb{H}$ grouping tests into equivalence classes. The hashing strategy depends on failure category:

\begin{itemize}[leftmargin=1em]
    \item \textit{For Exceptions ($\mathcal{T}_{exc}$):} We posit crashes occurring at the exact same stack trace location stem from the same defect. Let $S_t = [f_0, f_1, \dots, f_k]$ be the stack trace of test $t$, where $f_0$ is the top frame. We define:
    \[
    \Phi(t) = \mathcal{H}(\text{File}(f_0) \oplus \text{Func}(f_0) \oplus \text{Line}(f_0))
    \]
    
    \item \textit{For Assertion Failures ($\mathcal{T}_{asrt}$):} Since stack traces typically point to assertion utilities, we hash the AST node of the triggering statement to capture semantic check logic:
    \[
    \Phi(t) = \mathcal{H}(\text{Normalize}(\text{Stmt}_{trigger}(t)))
    \]
\end{itemize}
We select one representative test case for each unique hash value, yielding deduplicated set $\mathcal{T}_{dedup}$.

\textbf{(3) Validity Filtering.} Finally, we filter $\mathcal{T}_{dedup}$ to remove ``Invalid'' cases, tests failing due to incorrect test code rather than library bugs. This specifically targets $\mathcal{T}_{exc}$, as assertion failures inherently imply valid execution flow up to the check. We define a validity predicate $\mathcal{V}(t)$. Let $\mathcal{L}$ be the set of file paths belonging to the target library. A test $t$ is deemed \textbf{Invalid} if the exception is raised within the test script without engaging library logic:
\[
\neg \mathcal{V}(t) \iff (f_0 \in \text{TestFile}) \lor (\forall f_i \in S_t, f_i \notin \mathcal{L})
\]
Intuitively, a test is invalid if no stack frame appears within the target library (indicating failure is unrelated to the library), or the top stack frame originates from the test file itself (suggesting the exception is caused directly by test code). The final output is $\mathcal{T}_{unique} = \{ t \in \mathcal{T}_{dedup} \mid \mathcal{V}(t) \}$, serving as input for the Agentic Oracle. The complete deterministic process is outlined in \autoref{alg:preprocessing}.

\subsubsection{Agentic Oracle System}
\label{ssec:oracle}

Determining whether an anomaly constitutes a genuine defect requires deep semantic understanding of the framework's internal logic versus its documentation. By designing an oracle that actively explores the codebase, hypothesizes root causes, and verifies them against documentation through tool invocation, we can overcome knowledge obsolescence (C4), context saturation (C5), and passive observation bottlenecks (C6).

\begin{figure}[htbp]
    \centering
    \includegraphics[width=\linewidth]{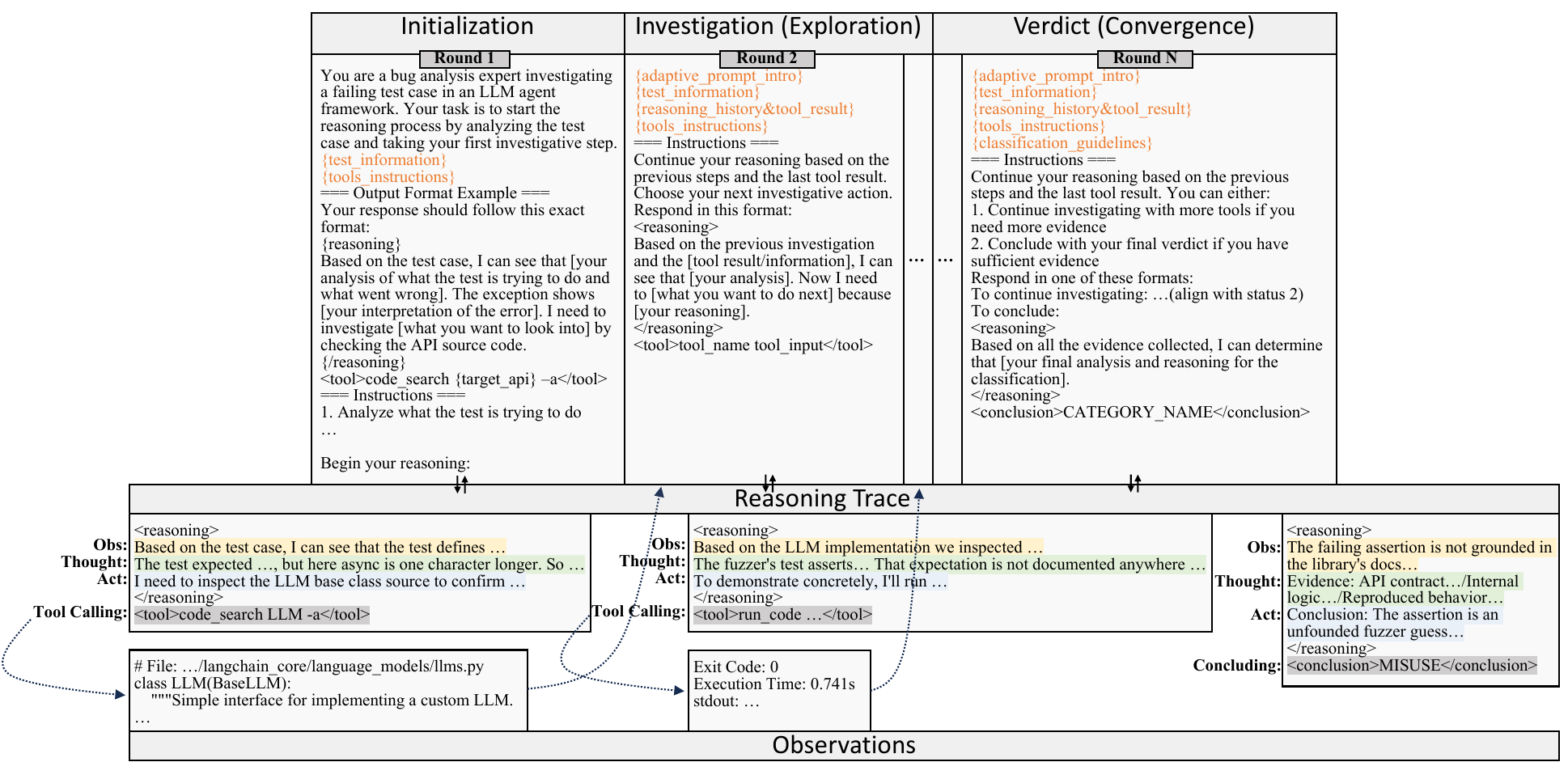}
    \caption{The workflow of the Agentic Oracle. The system utilizes Dual-Layer State Management to guide the agent through Initialization, Investigation, and Verdict phases. Concurrently, Dual-Stream Attention Management handles context by maintaining persistent reasoning traces while pruning observation windows.}
    \label{fig:agentic_oracle}
    \vspace{-1em}
\end{figure}

The candidate set filtered from the deterministic phase ($\mathcal{T}_{unique}$) represents high-potential anomalies. However, judging whether an anomaly is a genuine defect requires semantic analysis. We propose the \textbf{Agentic Oracle}, an automated triage system simulating expert debugging (\autoref{fig:agentic_oracle}). To enhance stability, we design a structured investigation loop with explicit control states. \textbf{Rather than operating as a passive classifier that outputs binary labels based on pre-determined context, our Agentic Oracle functions as an active reasoning agent}: it plans investigation strategies, queries relevant information sources (documentation, source code, runtime states), and verifies hypotheses through targeted tool use, mirroring how human developers debug complex failures.

\textbf{(1) Controller: Dual-Layer State Management (Addressing C4).} To ensure stable, evidence-driven reasoning rather than opaque one-shot judgments, the agent's control flow is managed by a dual-layer architecture: an inner ReAct~\cite{yao2023react} loop for step-by-step execution and an outer Finite State Machine (FSM)~\cite{stateflow} for global task progression.

\noindent\textit{\textbf{Inner Loop (ReAct Paradigm).}} Within each step, the agent follows the cycle of Observation $\rightarrow$ Reasoning $\rightarrow$ Action. It analyzes the current environment (results from the previous turn), formulates a plan, and executes tools or derives conclusions.

\noindent\textit{\textbf{Outer Loop (Explicit State Management).}} We divide each debugging session into three states to prevent aimless investigation:

\begin{itemize}[leftmargin=1em]
    \item \textbf{State 1: Initialization ($S_{init}$).} The agent analyzes test code, failure traceback, and specification documents to form a preliminary hypothesis.
    \item \textbf{State 2: Investigation ($S_{inv}$).} A mandatory exploration phase (rounds $< k_{min}$, where $k_{min}$ is a hyperparameter). The controller dynamically adjusts system prompts based on failure type (crash vs. assertion failure) to steer the agent toward legality checks or semantic expectation verification. In this phase, the agent employs tools to analyze API semantics, logic, and implementation to locate root causes, but is prohibited from rendering final verdicts.
    \item \textbf{State 3: Verdict ($S_{verdict}$).} Building on investigation, the agent receives guidance on error classification and is permitted to adjudicate root causes based on accumulated reasoning. The agent may continue using tools to locate or verify errors until the maximum round limit set by the controller is reached.
\end{itemize}

\textbf{(2) Memory: Dual-Stream Attention Management (Addressing C5).} To prevent attention drift in small-parameter models caused by excessive context, we partition information into two streams with differentiated retention strategies.

\noindent\textit{\textbf{Reasoning Trace (Persistent Memory).}} We maintain a compact reasoning trace across rounds to preserve the investigation context. This stream follows a fixed structure: \texttt{<reasoning> [Summary and analysis of previous tool result] -> [Current Hypothesis] -> [Next Plan] </reasoning>}.

\noindent\textit{\textbf{Observations (Pruned Memory).}} For large tool outputs such as logs or code snippets, we keep only a sliding observation window. Only observation results from the most recent round are retained in active context, ensuring the agent perceives immediate consequences of its actions without being overwhelmed by obsolete data.

\textbf{(3) Tooling: Introspection Capabilities (Addressing C4 and C6).} To enable active hypothesis verification against real program states rather than relying on parametric knowledge, the agent interacts with the environment through a specialized toolset for software analysis (\autoref{tab:oracle_tools}). \texttt{code\_search} uses Python reflection to locate the source of target objects such as classes, methods, and functions. It exposes a controlled ``jump-to-definition'' interface to the oracle, enabling precise source inspection while avoiding the safety and noise issues of unrestricted shell-based search. \texttt{run\_code} executes tests in an isolated sandbox with timeouts and fresh working directories.

\begin{table}[htbp]
\centering
\caption{Introspection tools for Agentic Oracle.}
\label{tab:oracle_tools}
\small  
\resizebox{\linewidth}{!}{
\begin{tabular}{lp{0.95\linewidth}}
\toprule
\textbf{Tool} & \textbf{Functionality} \\
\midrule
\texttt{code\_search} & Reflection-based jump-to-definition for locating target classes, methods, and functions. \\[0.5ex]  
\texttt{doc\_search} & Retrieves API documentation for verifying intended vs. actual behavior. \\[0.5ex]
\texttt{run\_code} & Executes reproduction scripts in sandbox to dynamically verify hypotheses. \\[0.5ex]
\texttt{get\_test\_info} & Obtains test failure details (return codes, execution time). Fallback tool for base information. \\
\bottomrule
\end{tabular}
}
\end{table}

\textbf{(4) Six-Dimensional Defect Taxonomy.} A critical challenge in LLM-based testing is the ``Overly Conservative Bias'', where models tend to label any unexpected behavior as a defect. To mitigate this, we define a fine-grained taxonomy aligned with ISO/IEC 25010 standards and API Misuse literature~\cite{misuse}, requiring the agent to map root causes into one of six categories (\autoref{tab:defect_taxonomy}).

\begin{table}[htbp]
\centering
\caption{Six-dimensional defect taxonomy for agent framework testing.}
\label{tab:defect_taxonomy}
\small  
\resizebox{\linewidth}{!}{
\begin{tabular}{@{}lp{0.6\linewidth}p{0.38\linewidth}@{}}
\toprule
\textbf{Category} & \textbf{Description} & \textbf{Diagnostic Focus} \\
\midrule
\textsc{Internal Error} & Faults in library implementation. & Locate fault line, rule out misuse. \\[0.5ex]
\textsc{Doc Mismatch} & Behavior contradicts documentation without crashing. & Verify documented vs. actual behavior. \\[0.5ex]
\textsc{Data Integrity} & Violates data integrity principles despite correct implementation (e.g., LangChain splitters losing info). & Check semantic correctness preservation. \\[0.5ex]
\textsc{Robustness} & Exceptions/errors on undefined inputs. & Assess defensive programming. \\[0.5ex]
\textsc{Usability} & Correct but counter-intuitive design. & Evaluate user-friendliness. \\[0.5ex]
\textsc{Misuse} & Wrong test expectations or unintended API usage. & Distinguish user vs. framework error. \\
\bottomrule
\end{tabular}
}
\vspace{-0.5em}
\end{table}

\noindent\textit{\textbf{Final Verdict Criteria.}} To ensure high value of reported defects, we classify \textsc{Internal Error}, \textsc{Doc Mismatch}, and \textsc{Data Integrity} as \textbf{Genuine Bugs}. Conversely, we categorize \textsc{Robustness}, \textsc{Usability}, and \textsc{Misuse} as \textbf{Non-Bugs}. Although \textsc{Robustness} and \textsc{Usability} represent valid improvement points, they are pervasive in software libraries; treating them as critical defects would dilute the significance of our findings and introduce noise common to traditional fuzzing. Finally, \textbf{to suppress stochastic errors in the LLM, we employ a High-Confidence Consensus (HCC) mechanism: a final alert is issued only if four independent Agentic Oracle sessions all classify the anomaly as a genuine bug.}

\section{Evaluation}
\label{sec:evaluation}

In this section, we conduct comprehensive experiments to address the following research questions:
\begin{itemize}[leftmargin=1em]
\item \textbf{RQ1 (Effectiveness):} Can \tool{} detect previously unknown bugs in real-world LLM agent frameworks?
\item \textbf{RQ2 (Test Case Generation Quality):} Compared to state-of-the-art baselines, can \tool{} generate higher-quality test cases that are more effective at exposing bugs?
\item \textbf{RQ3 (Oracle Accuracy and Cost):} How accurate and cost-effective is the Agentic Oracle compared to existing oracle approaches?
\item \textbf{RQ4 (Ablation Analysis):} How do the key components of \tool{} contribute to its overall effectiveness?
\end{itemize}

\subsection{Evaluation Setup}

\subsubsection{Environment}
Experiments were conducted on a Linux server running Ubuntu 22.04.3 LTS, equipped with an AMD EPYC 7713 Processor, 516GB of RAM, and two NVIDIA A100 GPUs.

\subsubsection{LLM Selection and Configuration}
\tool{} is implemented in Python. For the \textbf{Agentic Oracle} component and all oracle baselines evaluated in RQ3, we employ \textbf{GPT-5-mini} with temperature set to 0 to minimize hallucinations and ensure deterministic judgments. For the test generation phases, we use the \textbf{DeepSeek-V3} model with its default temperature settings, balancing code generation quality and cost efficiency. All LLM-based baselines in RQ2 also use DeepSeek-V3 to ensure fair comparison. To account for run-to-run variance in LLM-based oracle judgment, each LLM-based oracle baseline and the Agentic Oracle in RQ3 are repeated five times on the same oracle corpus. We report each metric as mean ± sample standard deviation across the five runs.

\subsubsection{Evaluation Targets}
To demonstrate the practical value of \tool{} in real-world scenarios, we evaluate on three widely-used open-source LLM agent frameworks: \textbf{LangChain}, \textbf{LlamaIndex}, and \textbf{CrewAI}. To the best of our knowledge, no prior automated testing research has been conducted on these frameworks, making them a representative and challenging testbed. Due to their large codebases, we focus on core libraries and APIs frequently used in actual open-source agent projects to maximize real-world relevance. Detailed statistics are provided in \autoref{tab:benchmarks}.

\begin{table}[ht]
\centering
\caption{Overview of the benchmark frameworks and our specific testing targets. Stars are recorded as of June 30, 2025.}
\label{tab:benchmarks}
\resizebox{0.8\linewidth}{!}{%
\begin{tabular}{rccccc}
\toprule
\textbf{Framework} & \textbf{Stars} & \textbf{Total LoC (Python)} & \textbf{Test Target(s)} & \textbf{APIs Covered} & \textbf{Version(s)} \\
\midrule
\textbf{CrewAI} & 33.5k & 34k & \texttt{crewai} & 46 & 0.130.0 \\
\textbf{LangChain} & 110k & 6.6M & \texttt{langchain-core} & 302 & 0.3.65 \\
\textbf{LlamaIndex} & 42.7k & 272k & \texttt{llama-index-core} & 264 & 0.12.42 \\
\bottomrule
\end{tabular}%
}
\end{table}

\subsubsection{Baselines}
We employ two distinct sets of baselines corresponding to the generation and oracle tasks.

\noindent\textbf{Test Generation Baselines.}
We compare \tool{} against four state-of-the-art tools:
\begin{itemize}[leftmargin=1em]
\item \textbf{Pynguin~\cite{pynguin}}: A state-of-the-art search-based test generation tool for Python. We use version 0.41.0 with its default DynaMOSA algorithm and 600-second search budget per target module.
\item \textbf{Fuzz4All~\cite{xia2024fuzz4all}}: A universal LLM-driven fuzzer using autoprompting. We adapted it by providing framework-specific documentation.
\item \textbf{TitanFuzz~\cite{deng2023titanfuzz}}: Originally designed for deep learning libraries, we adapted it to general Python libraries. TitanFuzz uses the initial seeds generated by \tool{} as its seed corpus.
\item \textbf{TELPA~\cite{telpa}}: A state-of-the-art LLM-based test generation tool combining large language models with program analysis. We evaluate only the Pynguin-based variant, as the CodaMOSA\cite{codamosa}-based variant is incompatible with all three target frameworks due to Pydantic version conflicts.
\end{itemize}

\noindent\textbf{Test Oracle Baselines.}
We compare our Agentic Oracle against passive approaches:
\begin{itemize}[leftmargin=1em]
\item \textbf{Raw Failure}: Treats any AssertionError or unhandled Exception as a bug.
\item \textbf{Heuristic Filter}: Applies our stack-trace-based filtering rules to exclude obvious test-level errors.
\item \textbf{Naive LLM Judge}: An LLM provided only with test code and failure traceback.
\item \textbf{LLM Judge w/ Doc}: An LLM with full relevant API documentation, representing a standard RAG-enhanced approach without active tool usage.
\end{itemize}
For the LLM-based baselines, we evaluate variants with both GPT-5-mini and the stronger GPT-5.2 backbone to reflect their best achievable performance. Both use the same Six-Dimensional Defect Taxonomy as the Agentic Oracle for fair comparison.

\subsubsection{Dataset Construction for Oracle Evaluation}
\label{sub:dataset_construction}
To evaluate test oracles under realistic fuzzing conditions, we construct a \textbf{Realistic Failure Corpus} of 1,000 unique failed test cases sampled from the raw generator outputs. All failures are deduplicated using stack-trace hashing.

From the unique failure pool, we apply proportional stratified sampling: \textbf{400} failures from LangChain, \textbf{400} from LlamaIndex, and \textbf{200} from CrewAI, reflecting their relative API surface and complexity.
Two developers independently labeled each failure according to the defect taxonomy defined in \autoref{ssec:oracle} and official documentation. Disagreements were resolved through consensus discussion, achieving a Cohen's Kappa of 0.77. The final corpus contains \textbf{28 genuine framework bugs} and \textbf{972 non-bug cases}, forming a highly imbalanced yet representative dataset for oracle evaluation.

In RQ3, we additionally use two supplementary datasets: a post-cutoff known-bug set with \textbf{31} reproducible real-world bugs mined from developer-fixed PRs, and a hard-negative set with \textbf{484} TitanFuzz-generated failures manually verified as non-bugs. Following \cite{xue_agentstudy}, we identify candidate bug-fixing PRs using keyword-based filtering, and retain only those created after GPT-5.2's training cutoff (2025-08-31) whose failures are reproducible on our selected versions using linked issue reproducers or PR regression tests. Since these datasets differ from the primary Realistic Failure Corpus in source and class distribution, we use them only as reference evaluations for known-bug recall and false-positive behavior.

\subsection{Evaluation Metrics}

\noindent\textbf{Generation Quality.}
\begin{itemize}[leftmargin=1.5em]
\item \textbf{Code Coverage}: Line coverage calculated on the source code of target API modules during execution.
\item \textbf{Valid Failed Test}: We classify generated tests into three hierarchical levels: (L1) \textit{Failed, Syntactically Valid Tests} that fail without syntax errors; (L2) \textit{Unique Failed Tests} after deduplication; (L3) \textit{Effective Failed Tests} that pass Validity filtering, representing genuinely valuable tests with high bug-finding potential.
\end{itemize}

\noindent\textbf{Oracle Effectiveness.}
We evaluate oracle accuracy using standard classification metrics: \textbf{Precision (P)}, \textbf{Recall (R)}, \textbf{F1}, and \textbf{False Positive Rate (FPR)}, computed against manually labeled outcomes. We define TP as genuine bugs correctly identified, FP as non-bug cases incorrectly flagged, FN as genuine bugs missed, and TN as non-bug cases correctly dismissed:
\[
\text{P}=\frac{TP}{TP+FP},\quad
\text{R}=\frac{TP}{TP+FN},\quad
\text{FPR}=\frac{FP}{FP+TN},\quad
\text{F1}=2\cdot\frac{\text{P}\cdot\text{R}}{\text{P}+\text{R}}
\]
In fuzzing triage, \textbf{Precision (P)} and \textbf{FPR} are especially critical, as false positives directly translate to unnecessary manual debugging cost. For the supplementary datasets in RQ3, we additionally report $R_{\mathrm{pos}}$, the recall on the post-cutoff known-bug set, and $\mathrm{FPR}_{\mathrm{neg}}$, the false positive rate on the hard-negative set.

\subsection{RQ1: Real-World Bug Detection}
\label{ssec:rq1}

\autoref{tab:bug_status} summarizes the bugs discovered by \tool{}. In total, \tool{} uncovered \textbf{40 unique and previously undisclosed bugs}, including \textbf{8 unexpected exception bugs} and \textbf{32 silent failure bugs}. Among them, \textbf{30 have been acknowledged by developers}, and \textbf{26 have already been fixed}. The count is based on unique developer-facing failure patterns after cross-test deduplication; the corresponding 32 root-cause clusters are reported separately in our artifact.

\begin{wraptable}{r}{0.45\linewidth}
\centering
\caption{Status of the reported bugs confirmed by developers.}
\label{tab:bug_status}
\vspace{-0.5em}
\resizebox{\linewidth}{!}{%
\begin{tabular}{rccc}
\toprule
\textbf{Framework} & \textbf{\# Total} & \textbf{\# Confirmed} & \textbf{\# Fixed} \\
\midrule
LlamaIndex & 19 & 16 & 15 \\ 
LangChain  & 18 & 12 & 10 \\
CrewAI     & 3 & 2 & 1 \\
\midrule
\textbf{Total} & \textbf{40} & \textbf{30} & \textbf{26} \\
\bottomrule
\end{tabular}%
}
\vspace{-0.5em}
\end{wraptable}

In contrast, \textbf{none of the baseline test generation tools identified any bugs under their original oracles in the three studied frameworks}. A closer inspection reveals the underlying reasons. Tools such as TitanFuzz and Fuzz4All rely on hard crashes (e.g., segmentation faults) as their primary oracle, while treating Python-level exceptions (e.g., \texttt{ValueError}, \texttt{TypeError}) as signals to filter out. However, since the target libraries are implemented entirely in Python, memory errors are extremely unlikely. All 8 unexpected-exception bugs we identified manifest as Python-level exceptions rather than memory faults.
For Pynguin and TELPA, the limitation stems from their design goals. These tools are not intended to discover bugs, but rather to safely increase code coverage for regression testing. As a result, most generated tests either pass successfully or are marked as expected failures. The few genuinely failing tests are almost exclusively caused by incorrect API usage in the test code rather than defects in the libraries themselves.

We further classified the discovered bugs using the taxonomy defined in \autoref{tab:defect_taxonomy}, yielding 15 Internal Errors, 10 Documentation Mismatches, and 15 Data Integrity Issues. Internal Errors typically arise from unhandled edge cases, leading to unexpected exceptions or behaviors inconsistent with intended semantics. Documentation Mismatches are mostly caused by incomplete implementations of documented functionality, with a smaller fraction involving violations of documented return-value contracts. Data Integrity Issues commonly result from flawed internal transformations, such as inadvertently converting empty values into \texttt{None}. While some overlap with Internal Errors, their undocumented silent transformations are particularly dangerous: even when no crash occurs, they can propagate downstream and cause latent failures. To clarify what kinds of bugs \tool{} discovers and why they matter, \autoref{tab:bug_component_impact} further groups the 40 bugs by their affected framework components. The largest group involves model and external-service integration (14 bugs), followed by workflow orchestration/callbacks and memory/retrieval/data stores (8 bugs each), showing that most bugs affect components directly involved in model I/O, workflow execution, and knowledge/context management.






\begin{table}
\centering
\caption{Component-level distribution of reported bugs.}
\label{tab:bug_component_impact}
\vspace{-0.5em}
\scriptsize
\setlength{\tabcolsep}{3pt}
\resizebox{\linewidth}{!}{%
\begin{tabular}{lcp{0.48\linewidth}}
\toprule
\textbf{Component} & \textbf{\# Bugs} & \textbf{Potential Impact} \\
\midrule
Workflow/callbacks 
& 8 
& Interrupted progression; inconsistent event state. \\

Memory/retrieval/storage 
& 8 
& Polluted context; incorrect ranking or provenance. \\

Schema/configuration 
& 6 
& Silent field loss; invalid serialization or config. \\

Model/external integration 
& 14 
& Failed calls; broken streaming or response conversion. \\

Tool/output parsing 
& 4 
& Missing tool calls; corrupted arguments or outputs. \\
\bottomrule
\end{tabular}%
}
\vspace{-0.8em}
\end{table}
A concrete example of these bugs will be presented in \autoref{sec:discusstions}, together with the actual reasoning process carried out by the Agentic Oracle.

\subsection{RQ2: Test Case Generation Quality}
\label{ssec:rq2}
\begin{table*}[htbp]
\centering
\small
\caption{Generation quality comparison (RQ2).}
\label{tab:rq2_generation}
\resizebox{0.8\textwidth}{!}{%
\begin{tabular}{rlrrrrr|rlrrrrr}
\toprule
\textbf{Method} & \textbf{Target} & \textbf{Total} & \textbf{L1} & \textbf{L2} & \textbf{L3} & \textbf{Cov.(\%)} 
& \textbf{Method} & \textbf{Target} & \textbf{Total} & \textbf{L1} & \textbf{L2} & \textbf{L3} & \textbf{Cov.(\%)} \\
\midrule
\multirow{3}{*}{\textbf{\tool{}}} 
& LangChain  & 20605 & 5401 & 3153 & 1557 & 61.95 
& \multirow{3}{*}{TitanFuzz} 
& LangChain  & 39435 & 26152 & 24300 & 352 & 54.22 \\
& LlamaIndex & 15242 & 4072 & 2436 & 1209 & 62.99 
& & LlamaIndex & 15810 & 10176 & 8842 & 123 & 56.96 \\
& CrewAI    & 4515 & 1708 & 989 & 493 & 69.24 
& & CrewAI    & 1202 & 732 & 668 & 10 & 50.36 \\
\midrule
\multirow{3}{*}{Fuzz4All} 
& LangChain  & 37290 & 26416 & 18521 & 8 & 19.33 
& \multirow{3}{*}{Pynguin} 
& LangChain  & 241 & 5 & 5 & 1 & 45.96 \\
& LlamaIndex & 24420 & 16217 & 15285 & 4 & 31.90 
& & LlamaIndex & 167 & 28 & 17 & 1 & 52.59 \\
& CrewAI    & 13020 & 8342 & 5965 & 6 & 25.11 
& & CrewAI    & 79 & 4 & 1 & 1 & 38.15 \\
\midrule
\multirow{3}{*}{TELPA} 
& LangChain  & 999 & 12 & 12 & 0 & 45.99 
& & & & & & & \\
& LlamaIndex & 529 & 17 & 17 & 0 & 50.52 
& & & & & & & \\
& CrewAI    & 548 & 27 & 27 & 0 & 38.97 
& & & & & & & \\
\bottomrule
\end{tabular}%
}
\vspace{-1em}
\end{table*}

We evaluate different automated test generation approaches from two perspectives: code coverage and the quality of failing test cases. \tool{} consistently generates a substantial number of structurally complex test cases across all three target frameworks, achieving overall higher or significantly better code coverage than existing LLM-based fuzzing and search-based tools (\autoref{tab:rq2_generation}).

We further analyze the factors limiting baseline coverage. TitanFuzz and FuzzGPT perform well on simple functions or standalone class methods, but struggle with class-level APIs or SPIs. SBST-based tools such as Pynguin and TELPA are constrained by their design and fail to exercise the asynchronous methods that are pervasive in all three frameworks, leading to limited coverage.

We examine whether the generated tests are valuable for use with the Agentic Oracle. Judging solely by the number of failing tests, traditional fuzzing tools (e.g., TitanFuzz and Fuzz4All) appear more aggressive, producing a large number of failures at the L1 and L2 levels. However, these failures are predominantly caused by invalid inputs, repetitive patterns, or shallow API misuse. After deduplication and effectiveness filtering, the number of L3 valid failing tests drops sharply to single digits for some targets, indicating an extremely high noise level that makes them difficult to translate into real defect discovery. We further cross-evaluated all baseline-generated L3 failures using the Agentic Oracle and manual verification. Among 506 such failures, only one genuine bug was found, from TitanFuzz, and it corresponds to LangChain Issue~\#30640, which was also discovered by \tool{}. This indicates that the baselines are limited not only by weak oracles, but also by the low bug density of their generated failures.
In contrast, \tool{} exhibits a substantially higher failure density throughout the progressive filtering from L1 to L3. Although its overall generation scale is smaller than some baselines, it produces the largest number of L3 valid failing test cases on all three target systems, significantly outperforming all competitors. This demonstrates that the failing tests generated by \tool{} are not merely byproducts of random perturbations, but instead systematically probe semantic boundaries and implicit constraints of the APIs.

The performance of Pynguin and TELPA reflects a different trade-off. As these tools are primarily designed for safely increasing coverage and supporting regression testing, they generate relatively few test cases overall. Moreover, most failures are filtered out early as expected failures or invalid tests, resulting in almost no effective failing tests. This further illustrates that optimizing solely for coverage is insufficient to support real-world bug discovery.

\subsection{RQ3: Oracle Accuracy and Cost}
\label{ssec:rq3}
\begin{table}[h]
\centering
\caption{Oracle accuracy, cost, and supplementary-set comparison (RQ3).}
\label{tab:rq3_oracle}
\resizebox{\linewidth}{!}{%
\begin{tabular}{rccccccccc}
\toprule
\textbf{Oracle} 
& \textbf{P(\%)} 
& \textbf{R(\%)} 
& \textbf{F1(\%)} 
& \textbf{FPR(\%)} 
& \textbf{Cost(\$)} 
& \textbf{\$/Bug} 
& \textbf{Reviews/Bug}
& \textbf{$R_{\mathrm{pos}}$(\%)}
& \textbf{$\mathrm{FPR}_{\mathrm{neg}}$(\%)} \\
\midrule
Raw Failure (Base Oracle) 
& 2.80 & 100.00 & 5.45 & 100.00
 & -- & -- & 35.71 
& 100.00 & 100.00 \\

Heuristic Filter 
& 5.82 & 100.00 & 11.00 &46.60
& -- & -- & 17.24 
& 100.00 & 100.00 \\

Naïve LLM Judge 
& 12.40±0.63 & 94.64±2.06 & 21.92±1.02 & 19.29±0.84 
& 0.54 & 0.02 & 8.06 
& 93.55±2.28 & 17.52±1.50 \\

LLM Judge w/ Doc 
& 12.78±0.22 & 88.57±2.99 & 22.34±0.39 & 17.41±0.50 
& 0.58 & 0.02 & 7.82 
& 94.84±1.77 & 8.39±0.63 \\

Naïve LLM Judge (GPT-5.2) 
& 29.27±5.40 & 59.29±4.07 & 39.01±4.99 & 4.24±0.84 
& 3.21 & 0.19 & 3.42 
& 70.32±3.53 & 2.07±0.41 \\

LLM Judge w/ Doc (GPT-5.2) 
& 29.02±4.81 & 65.71±5.98 & 39.97±4.16 & 4.77±1.10 
& 3.24 & 0.18 & 3.45 
& 67.74±3.23 & 1.03±0.29 \\

\textbf{Agentic Oracle (within \tool{})} 
& \textbf{91.17±4.01} & \textbf{72.14±1.60} & \textbf{80.49±1.34} & \textbf{0.21±0.10} 
& 5.10 & 0.25 & 1.10 
& 60.65±1.44 & 0.04±0.09 \\

\bottomrule
\end{tabular}%
}

\end{table}

\autoref{tab:rq3_oracle} reports the accuracy and cost comparison between the proposed Agentic Oracle and representative passive oracle baselines.

We first observe that implicit oracles are ineffective for agent frameworks. The Raw Failure oracle, which treats any failure as a bug, captures all bugs by construction (R = 100.00\%), but suffers from extremely low precision (2.80\%) and a 100.00\% false positive rate, making it impractical in real settings. The Heuristic Filter reduces part of the noise by excluding obvious test-level errors, yet still achieves only 5.82\% precision and requires 17.24 manual reviews to identify a single real bug. These results confirm that raw failure signals and simple heuristics cannot reliably distinguish genuine framework defects from invalid test cases.

Passive LLM-based judges improve over implicit oracles, but remain limited. With GPT-5-mini, the Naïve LLM Judge and LLM Judge w/ Doc achieve only 12.40\%$\pm$0.63\% and 12.78\%$\pm$0.22\% precision, respectively, while their FPRs remain high at 19.29\%$\pm$0.84\% and 17.41\%$\pm$0.50\%. Upgrading the backbone to GPT-5.2 improves precision to around 29\% and reduces FPR to around 4\%--5\%, but recall remains only 59.29\%$\pm$4.07\% and 65.71\%$\pm$5.98\%. This exposes a clear trade-off: stronger passive judges are more selective, but still suffer from non-negligible false positives and unstable recall.

In contrast, the Agentic Oracle reaches a qualitatively different operating point. It achieves 91.17\%$\pm$4.01\% precision, 72.14\%$\pm$1.60\% recall, 80.49\%$\pm$1.34\% F1, and only 0.21\%$\pm$0.10\% FPR. The small standard deviations show that the result is not driven by a single favorable run. Compared with GPT-5.2-based passive judges, the Agentic Oracle exhibits lower variance in recall, F1, and FPR, while also achieving much higher precision. Although its total monetary cost is slightly higher, the cost per discovered bug remains comparable, and the human auditing burden is dramatically reduced: only 1.10 failed test cases need to be reviewed to confirm a real bug, compared with 3.42--8.06 for LLM-based passive judges. In practice, industrial bug detection systems typically require precision above 90\% to be deployable in automated pipelines~\cite{tricorder}. Under this criterion, the Agentic Oracle is the only approach that meets deployment-level reliability.

We also evaluate the oracle on the supplementary datasets introduced in \S4.1. The post-cutoff known-bug set comes from real-world developer-fixed bugs rather than failures generated by testing tools, so we use it only as a reference indicator of external validity rather than as the primary oracle benchmark. On this set, the Agentic Oracle achieves 60.65\%$\pm$1.44\% recall, showing that it can still detect a majority of real-world bugs outside \tool{}'s own generated failures. Its recall is lower than that of passive LLM judges because the Agentic Oracle requires stronger evidence from source-code inspection and executable validation before reporting a bug, and some real-world bugs require complex system construction, external-service interaction, or multi-step agent workflows to fully confirm.

The hard-negative set further highlights the Agentic Oracle's conservative behavior. It achieves an FPR of only 0.04\%$\pm$0.09\%, which is about 1/25 of the best passive GPT-5.2 baseline on the same set. This result is particularly important for our setting: the Agentic Oracle is used as a post-processing oracle for noisy fuzzing outputs, where false positives dominate manual cost. Thus, the supplementary results show that the Agentic Oracle preserves strong false-positive control beyond the primary benchmark.

We also separately examined probe-triggered failures to distinguish probe activation from final oracle judgment. Raw Failure represents the unvalidated probe-signal baseline, where probe-triggered assertion failures are directly treated as bug reports. On the same probe-triggered subset, the Agentic Oracle achieves 91.55\%$\pm$3.61\% precision, 80.00\%$\pm$1.86\% recall, 85.35\%$\pm$1.80\% F1, and 0.45\%$\pm$0.21\% FPR. This supports our design: probes surface suspicious silent failures, while the Agentic Oracle validates them before issuing bug reports.

\begin{wrapfigure}{r}{0.45\linewidth}
    \centering
    \includegraphics[width=\linewidth]{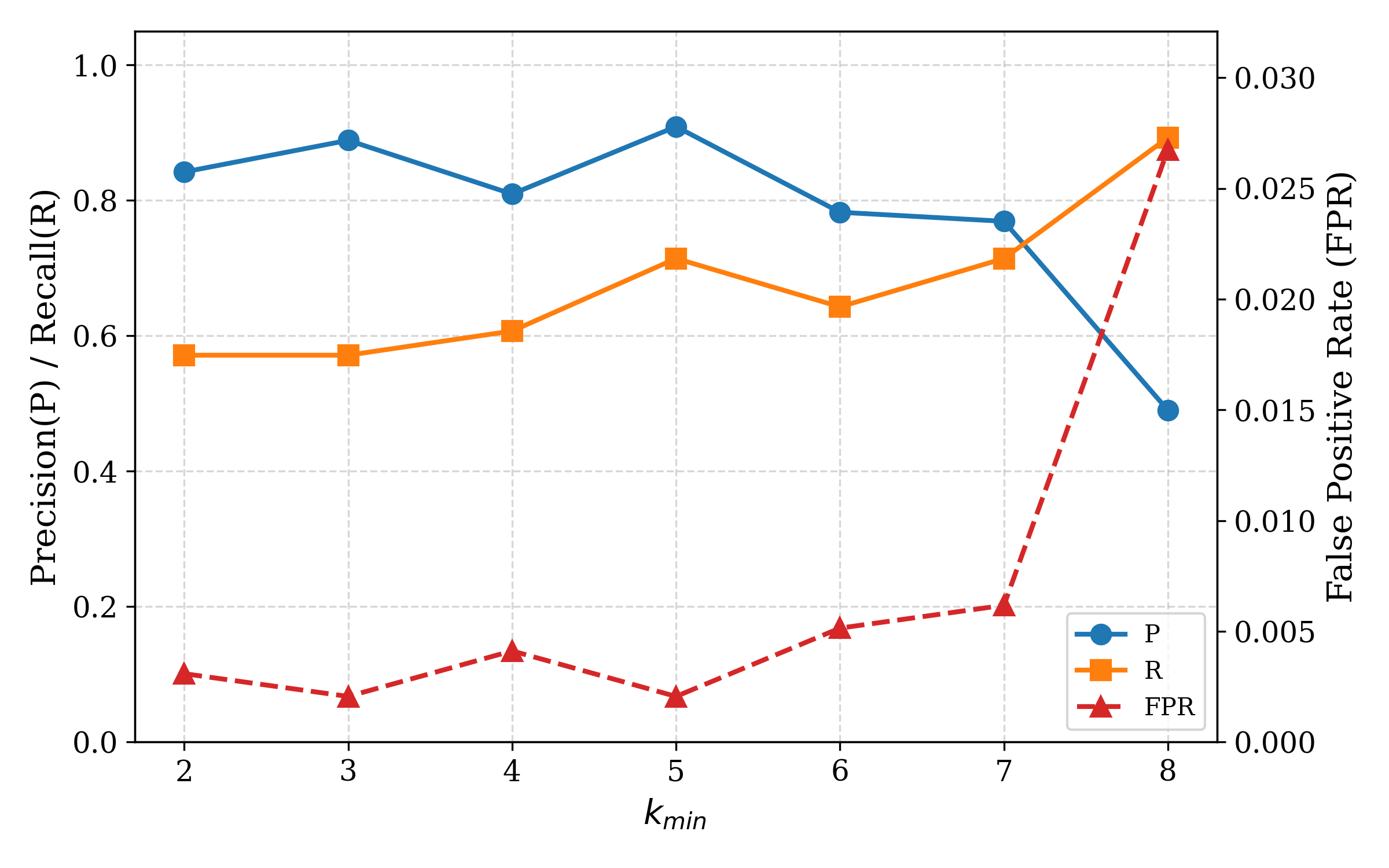}
    \caption{Impact of $k_{min}$ on oracle performance.}
    \label{fig:plot}
\end{wrapfigure}

\subsection{RQ4: Ablation Analysis}
\label{ssec:rq4}

We conducted a systematic ablation study on the key components and parameters of \tool{}, covering both the generation side and the oracle side: usage mining in seed synthesis, the state management strategy (minimum enforced reasoning rounds $k_{min}$) (\autoref{fig:plot}), the agent architecture, and the tool usage mechanism (\autoref{tab:rq4_ablation}).

For $k_{min}$, we investigated its effect when set to values ranging from 2 to the system's maximum allowed reasoning rounds ($k_{max}$=8), with the default value being 5. For the tools ablation, we evaluated the following variants: \texttt{no\_tools} disables all tools except the base tool (\texttt{get\_test\_information}), meaning the agent can only access information from the test case itself; \texttt{no\_run\_code} disables \texttt{run\_code}, so the agent can only investigate through static analysis; \texttt{no\_code\_search} disables \texttt{code\_search}, preventing direct inspection of API source code; and \texttt{no\_spec} disables both \texttt{code\_search} and \texttt{doc\_search}, leaving the agent with no direct access to external information. For the architecture ablations, we evaluate three variants. The \texttt{full} variant replaces our Dual-Stream Attention Management with a simple full-context strategy that provides the LLM with all previous reasoning outputs and tool call results without pruning. The Vanilla ReAct Agent uses the same evidence-gathering tools but removes the structured state machine, dual-memory management, and HCC mechanism. The \texttt{w/o HCC} variant preserves the Agentic Oracle workflow but removes the high-confidence consensus check before issuing a bug verdict.

Regarding state management, we observe that as $k_{min}$ increases, precision generally declines while recall rises, reaching an optimal balance around round 5. Lower $k_{min}$ values lead to insufficient reasoning depth, causing the agent to make premature judgments with inadequate evidence, which results in low recall due to our conservative strategy settings. Conversely, as $k_{min}$ approaches the system's maximum limit, reasoning is forcibly prolonged; to satisfy the $k_{min}$ requirement, the agent searches for non-existent clues, inducing hallucinations, while excessive context interferes with attention allocation~\cite{liu2023lost}, leading to degraded precision. Notably, at $k_{min}=8$, precision drops and FPR rises significantly. Since round 8 is the set maximum, the agent is forced to adjudicate immediately after receiving classification guidelines without the opportunity for further tool-based verification, confirming the necessity of the Verdict phase. 





\begin{wraptable}{hr}{0.65\linewidth}
\centering
\caption{Ablation study of Agentic Oracle components (RQ4).}
\label{tab:rq4_ablation}
\resizebox{0.85\linewidth}{!}{%
\begin{tabular}{lcccc}
\toprule
\textbf{Configuration} & \textbf{P(\%)} & \textbf{R(\%)} & \textbf{F1(\%)} & \textbf{FPR(\%)} \\
\midrule
\multicolumn{5}{l}{\textit{Architecture}} \\
Vanilla ReAct Agent 
& 54.55 & 85.71 & 66.67 & 2.06 \\

w/o HCC 
& 66.67 & 85.71 & 75.00 & 1.23 \\

w/o Dual-Memory (Full Context) 
& 70.83 & 60.71 & 65.38 & 0.72 \\

\textbf{Default Agentic Oracle} 
& \textbf{90.91} & \textbf{71.43} & \textbf{80.00} & \textbf{0.21} \\

\midrule
\multicolumn{5}{l}{\textit{Tool Usage}} \\
no\_tools 
& 90.91 & 35.71 & 51.28 & 0.10 \\

no\_run\_code 
& 82.61 & 67.86 & 74.51 & 0.41 \\

no\_code\_search 
& 81.82 & 64.29 & 72.00 & 0.41 \\

no\_spec 
& 65.52 & 67.86 & 66.67 & 1.03 \\

\textbf{Default (All Tools)} 
& \textbf{90.91} & \textbf{71.43} & \textbf{80.00} & \textbf{0.21} \\
\bottomrule
\end{tabular}%
}
\vspace{-0.5em}
\end{wraptable}
The tool ablation further validates the importance of multi-source external capabilities for agent discrimination. As shown in \autoref{tab:rq4_ablation}, completely disabling all tools except the base tool maintains high precision but causes a precipitous drop in recall, indicating that without external information, the agent can only identify a few high-confidence anomalies and fails to cover hidden logical deviations. In contrast, disabling \texttt{run\_code} or \texttt{code\_search} individually causes relatively limited performance degradation. We found this is because the agent autonomously seeks alternatives in the absence of specific tools (e.g., using Python reflection within \texttt{run\_code} to analyze source code when \texttt{code\_search} is disabled), demonstrating system robustness. However, when the ability to directly retrieve external information is removed (\texttt{no\_spec}), all metrics degrade significantly. In this scenario, the agent is forced to rely solely on code execution, an inefficient and error-prone method, to gather information, proving the necessity of external retrieval tools.

The architecture ablation shows that tool access alone is insufficient. The Vanilla ReAct Agent achieves high recall (85.71\%), but its precision drops to 54.55\% and FPR rises to 2.06\%, indicating that an unconstrained tool-using loop tends to over-report suspicious failures. Removing HCC exhibits a similar pattern: recall remains high at 85.71\%, but precision drops from 90.91\% to 66.67\% and FPR increases from 0.21\% to 1.23\%, confirming the role of HCC in suppressing premature bug reports and enforcing high-confidence decisions. The \texttt{w/o Dual-Memory} variant further shows that simply retaining the full context does not improve discrimination; instead, it reduces all metrics by introducing attention dispersion and noise accumulation. Collectively, these results show that the Agentic Oracle's performance depends on the joint effect of structured state control, high-confidence consensus, and pruned memory management.

\begin{table}[ht]
\centering
\caption{Ablation on usage mining for seed synthesis (RQ4).}
\label{tab:rq4_usage_mining}
\resizebox{0.78\linewidth}{!}{%
\begin{tabular}{lccc}
\toprule
\textbf{Metric} & \textbf{w/ Usage Mining} & \textbf{w/o Usage Mining} & \textbf{$\Delta$} \\
\midrule
Seed coverage (\%) 
& 54.68 & 52.33 & +2.35 \\

Raw seed pass rate (\%) 
& 62.55 & 64.28 & -1.73 \\

Post-fix pass rate (\%) 
& 78.40 & 80.84 & -2.44 \\

Final valid seeds 
& 4,324 & 2,074 & +108.5\% \\
\bottomrule
\end{tabular}%
}
\vspace{-0.5em}
\end{table}

We further ablate usage mining in seed synthesis. \autoref{tab:rq4_usage_mining} shows that usage mining slightly increases seed coverage and more than doubles the final valid seed count, from 2,074 to 4,324. Although the raw pass rate is slightly lower, this is expected because mined examples introduce more realistic and complex object orchestration contexts. The result also reflects the importance of the Fixer Agent: under the usage-mining setting, which improves the seed pass rate by 15.85\% on average by repairing execution failures in these complex seeds.
\section{Discussion}
\label{sec:discusstions}

\subsection{Case Study}
\label{ssec:case_study}

\begin{wrapfigure}{r}{0.6\linewidth}
\vspace{-1.2em}
\centering
\includegraphics[width=\linewidth, trim=60 170 500 140, clip]{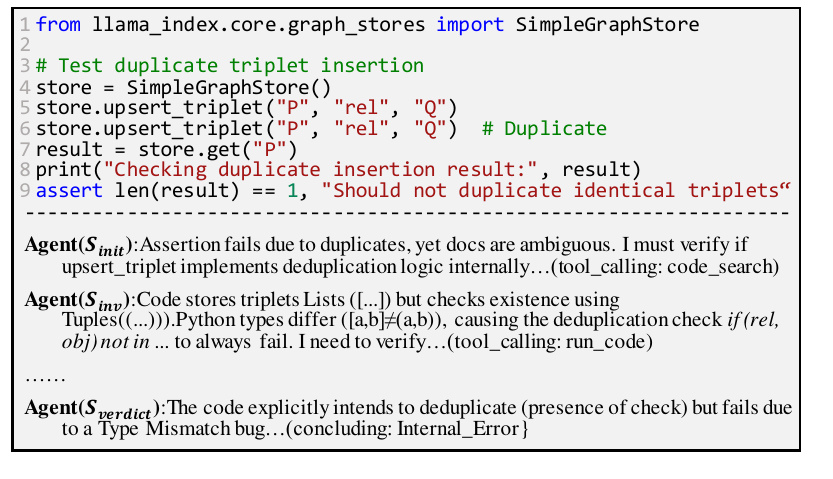}
\caption{Case study of an internal deduplication bug in LlamaIndex.}
\label{fig:case_study_code}
\vspace{-1.2em}
\end{wrapfigure}



We present a case study of a bug discovered by \tool{} (\autoref{fig:case_study_code}). 
The bug breaks the idempotency of \texttt{SimpleGraphStore.\allowbreak upsert\_triplet} in 
LlamaIndex, allowing duplicate triplets to be inserted silently. Since the execution 
does not raise an exception, the defect is exposed only by the behavioral probe 
synthesized by \tool{} (line 9), which checks that repeated \texttt{upsert} calls 
preserve a single logical relation.

The Agentic Oracle localizes the root cause with \texttt{code\_search}: triplets are 
stored as lists, while duplicate checks compare against tuples, so semantically 
identical triplets are treated as different objects. This conventional implementation 
defect can affect downstream agent infrastructure. In a real-world agent system we 
inspected, the framework graph store is used to build a tool knowledge graph for 
planning and tool selection; extracted triplets are repeatedly inserted through 
\texttt{KnowledgeGraphIndex.upsert\_triplet} and persisted with 
\texttt{SimpleGraphStore}. Once deduplication silently fails, identical relations can 
be persisted multiple times, inflating the knowledge graph and biasing retrieval or 
planning toward repeated evidence. The explicit reasoning trace produced by the 
Agentic Oracle made manual inspection straightforward, and the issue was promptly 
fixed after reporting.

\subsection{Limitations and Future Work}
\label{sec:limitations}

\tool{} has several limitations. First, generation quality depends on high-quality real-world usage examples, so seeds may degrade for novel or sparsely used APIs. Second, its API-centric design is less suited to bugs requiring complex multi-API interactions or external environmental states, such as database contents or live web service responses. Third, although the Agentic Oracle has reasonable monetary cost, multi-round LLM reasoning limits latency, and nondeterminism cannot be fully eliminated despite temperature-zero decoding and HCC. Finally, our evaluation uses DeepSeek-V3 for generation, GPT-5-mini for oracle diagnosis, and three Python-based LLM agent frameworks. While the methodology can be extended to Python frameworks such as FastAPI\cite{Ramirez_FastAPI} and, in principle, to other languages and ecosystems, broader cross-model, cross-framework, and cross-language validation remains future work.

\section{Related Work}
\label{sec:related work}

\textbf{LLM Agent Framework.} Recent research explores LLM Agent Frameworks.
One line of work focuses on \textbf{performance and scalability}, benchmarking popular frameworks like LangChain\cite{Chase_LangChain_2022} and LlamaIndex\cite{Liu_LlamaIndex_2022} to guide enterprise adoption~\cite{benchmark_}.
Another area is \textbf{security}. 
Studies investigate vulnerabilities arising from the interaction with external tools~\cite{milev2025toolfuzzautomatedagent, li2025Les_2025}, as well as framework-level risks such as Remote Code Execution (RCE) and command injection~\cite{RCE_2024, wang2025leveraginglargelanguagemodels}.
A third area explores the \textbf{maintainability challenges} of these systems, with work like~\cite{rahardja2025agentsfixagentissues} constructing benchmarks to evaluate the difficulty of automatically resolving issues in agent-based software.
Our work complements these by focusing on a different dimension: the automated discovery of general-purpose logic bugs within the core APIs of the frameworks themselves.

\noindent\textbf{LLMs for Software Testing.} In the era of LLMs, an increasing number of approaches use LLMs for automated test generation, including unit testing~\cite{chattester,codamosa,telpa,chen2024chatunitest} and fuzz testing~\cite{deng2023titanfuzz,Deng2024_FuzzGPT,xia2024fuzz4all,zhang2025DFuzz,Ou2026dlfuzz}. With the advent of LLMs, people have also begun using them for automated assertion generation~\cite{he2024empirical, aid2024, doc2oracle2024, hayet2025chatassert}. CANDOR~\cite{candor2024} and Nexus~\cite{nexas} go further by employing a multi-agent framework to generate more accurate oracles. These automated assertion generation methods fundamentally differ from our work in the type of oracle problem they address: they target \emph{functional verification}, producing prescriptive outputs to check whether an implementation meets its specification, whereas our Agentic Oracle targets \emph{bug discovery}, adjudicating whether anomalous executions represent genuine semantic defects in an undefined or ambiguous input space.

\section{Conclusion}
\label{sec:conclusion}

We introduce \tool{}, bridging the gap between specification-driven generation and active defect diagnosis via an Agentic Oracle. By orchestrating autonomous documentation retrieval and runtime introspection, \tool{} overcomes the limitations of passive LLM-based verification. In our evaluation, \tool{} discovered 40 unknown bugs (30 confirmed, 26 fixed), while state-of-the-art baselines identified none. Achieving 91.17\% precision, a 61 percentage point increase over the best passive baseline, \tool{} establishes a highly accurate paradigm for securing the foundational infrastructure of modern AI agents.

\section*{Acknowledgements}
This work was supported in part by the National Natural Science Foundation of China (grants No. 62572209, 62502168), and by the Hubei Provincial Key Research and Development Program (grant No. 2025BAB057).

We acknowledge the use of AI as a writing assistance tool. These AI systems were employed exclusively for sentence polishing, grammar correction, and readability improvement. All technical contributions, including research design, implementation, experiments, analysis, and conclusions, were developed solely by the authors. AI tools were not used for generating figures, tables, code, or substantive content.

\newpage
\section*{Data Availability}
Our artifact is available at \url{https://github.com/security-pride/LogicHunter}.

\bibliographystyle{ACM-Reference-Format}
\bibliography{main}

@inproceedings{deng2023titanfuzz,
  title={{Large Language Models are Zero-Shot Fuzzers: Fuzzing Deep-Learning Libraries via Large Language Models}},
  author={Deng, Y. and Xia, C. S. and Peng, H. and Yang, C. and Zhang, L.},
  booktitle={Proceedings of the 32nd ACM International Symposium on Software Testing and Analysis (ISSTA 2023)},
  pages={423--435},
  year={2023}
}

@inproceedings{zhang2025DFuzz,
  title={{Your Fix Is My Exploit: Enabling Comprehensive DL Library API Fuzzing with Large Language Models}},
  author={Zhang, K. and Wang, S. and Han, J. and Zhu, X. and Li, X. and Wang, S. and others},
  booktitle={2025 IEEE/ACM 47th International Conference on Software Engineering (ICSE)},
  pages={508--520},
  year={2025}
}

@article{llm4se,
  title={{Large Language Models for Software Engineering: A Systematic Literature Review}},
  author={Hou, X. and Zhao, Y. and Liu, Y. and Yang, Z. and Wang, K. and Li, L. and others},
  journal={ACM Transactions on Software Engineering and Methodology},
  volume={33},
  number={8},
  pages={220},
  year={2024}
}

@article{oracleinllm,
  title={{Test Oracle Automation in the Era of LLMs}},
  author={Molina, F. and Gorla, A. and d'Amorim, M.},
  journal={ACM Transactions on Software Engineering and Methodology},
  volume={34},
  number={1},
  pages={1},
  year={2025}
}

@article{hayet2025chatassert,
  title={{ChatAssert: LLM-Based Test Oracle Generation With External Tools Assistance}},
  author={Hayet, I. and Scott, A. and d'Amorim, M.},
  journal={IEEE Transactions on Software Engineering},
  volume={51},
  number={1},
  pages={305--319},
  year={2025}
}

@inproceedings{li2023pyrtfuzz,
  title={{PyRTFuzz: Detecting Bugs in Python Runtimes via Two-Level Collaborative Fuzzing}},
  author={Li, W. and Yang, H. and Luo, X. and Cheng, L. and Cai, H.},
  booktitle={Proceedings of the 2023 ACM SIGSAC Conference on Computer and Communications Security},
  pages={1645--1659},
  year={2023}
}

@misc{konstantinou2024oracles,
  title={{Do LLMs generate test oracles that capture the actual or the expected program behaviour?}},
  author={Konstantinou, M. and Degiovanni, R. and Papadakis, M.},
  year={2024},
  eprint={2410.21136},
  archivePrefix={arXiv}
}

@inproceedings{xia2024fuzz4all,
  title={{Fuzz4ALL: Universal Fuzzing with Large Language Models}},
  author={Xia, C. S. and Paltenghi, M. and Tian, J. L. and Pradel, M. and Zhang, L.},
  booktitle={2024 IEEE/ACM 46th International Conference on Software Engineering (ICSE)},
  pages={1547--1559},
  year={2024}
}

@article{he2024empirical,
  title={{An Empirical Study on Focal Methods in Deep-Learning-Based Approaches for Assertion Generation}},
  author={He, Y. and Huang, J. and Yu, H. and Xie, T.},
  journal={Proceedings of the ACM on Software Engineering},
  volume={1},
  number={FSE},
  pages={1750--1771},
  year={2024}
}

@inproceedings{chen2024chatunitest,
  title={{ChatUniTest: A Framework for LLM-Based Test Generation}},
  author={Chen, Y. and Hu, Z. and Zhi, C. and Han, J. and Deng, S. and Yin, J.},
  booktitle={Companion Proceedings of the 32nd ACM International Conference on the Foundations of Software Engineering},
  pages={572--576},
  year={2024}
}

@software{Chase_LangChain_2022,
author = {Chase, Harrison},
month = oct,
title = {{LangChain}},
url = {https://github.com/langchain-ai/langchain},
year = {2022}
}

@software{Liu_LlamaIndex_2022,
author = {Liu, Jerry},
month = {11},
title = {{LlamaIndex}},
url = {https://github.com/jerryjliu/llama_index},
year = {2022}
}

@misc{crewai2025,
  author = {CrewAI Inc.},
  title = {CrewAI: A lean, lightning-fast Python framework for autonomous AI agents},
  year = {2025},
  howpublished = {\url{https://github.com/crewAIInc/crewAI}},
  note = {Accessed: 2025-07-07}
}

@software{Colvin_Pydantic_2025,
author = {Colvin, Samuel and Jolibois, Eric and Ramezani, Hasan and Garcia Badaracco, Adrian and Dorsey, Terrence and Montague, David and Matveenko, Serge and Trylesinski, Marcelo and Runkle, Sydney and Hewitt, David and Hall, Alex and Plot, Victorien},
license = {MIT},
month = jun,
title = {{Pydantic}},
url = {https://github.com/pydantic/pydantic},
version = {v2.11.7},
year = {2025}
}

@misc{atheris2020,
  author       = {Google},
  title        = {Atheris: Coverage‑Guided Python Fuzzing Engine},
  howpublished = {\url{https://github.com/google/atheris}},
  note         = {Accessed: 2025-07-07},
}

@inproceedings{Deng2024_FuzzGPT,
author = {Deng, Yinlin and Xia, Chunqiu Steven and Yang, Chenyuan and Zhang, Shizhuo Dylan and Yang, Shujing and Zhang, Lingming},
title = {Large Language Models are Edge-Case Generators: Crafting Unusual Programs for Fuzzing Deep Learning Libraries},
year = {2024},
isbn = {9798400702174},
publisher = {Association for Computing Machinery},
address = {New York, NY, USA},
url = {https://doi.org/10.1145/3597503.3623343},
doi = {10.1145/3597503.3623343},
booktitle = {Proceedings of the IEEE/ACM 46th International Conference on Software Engineering},
articleno = {70},
numpages = {13},
location = {Lisbon, Portugal},
series = {ICSE '24}
}

@misc{milev2025toolfuzzautomatedagent,
      title={ToolFuzz -- Automated Agent Tool Testing}, 
      author={Ivan Milev and Mislav Balunović and Maximilian Baader and Martin Vechev},
      year={2025},
      eprint={2503.04479},
      archivePrefix={arXiv},
      primaryClass={cs.AI},
      url={https://arxiv.org/abs/2503.04479}, 
}

@misc{wang2025leveraginglargelanguagemodels,
      title={Leveraging Large Language Models for Command Injection Vulnerability Analysis in Python: An Empirical Study on Popular Open-Source Projects}, 
      author={Yuxuan Wang and Jingshu Chen and Qingyang Wang},
      year={2025},
      eprint={2505.15088},
      archivePrefix={arXiv},
      primaryClass={cs.SE},
      url={https://arxiv.org/abs/2505.15088}, 
}

@misc{rahardja2025agentsfixagentissues,
      title={Can Agents Fix Agent Issues?}, 
      author={Alfin Wijaya Rahardja and Junwei Liu and Weitong Chen and Zhenpeng Chen and Yiling Lou},
      year={2025},
      eprint={2505.20749},
      archivePrefix={arXiv},
      primaryClass={cs.AI},
      url={https://arxiv.org/abs/2505.20749}, 
}

@inproceedings{RCE_2024, series={CCS ’24},
   title={Demystifying RCE Vulnerabilities in LLM-Integrated Apps},
   url={http://dx.doi.org/10.1145/3658644.3690338},
   DOI={10.1145/3658644.3690338},
   booktitle={Proceedings of the 2024 on ACM SIGSAC Conference on Computer and Communications Security},
   publisher={ACM},
   author={Liu, Tong and Deng, Zizhuang and Meng, Guozhu and Li, Yuekang and Chen, Kai},
   year={2024},
   month=dec, pages={1716–1730},
   collection={CCS ’24} }

@misc{li2025Les_2025,
      title={Les Dissonances: Cross-Tool Harvesting and Polluting in Multi-Tool Empowered LLM Agents}, 
      author={Zichuan Li and Jian Cui and Xiaojing Liao and Luyi Xing},
      year={2025},
      eprint={2504.03111},
      archivePrefix={arXiv},
      primaryClass={cs.CR},
      url={https://arxiv.org/abs/2504.03111}, 
}

@inproceedings{Pynguin, series={ICSE ’22},
   title={Pynguin: automated unit test generation for Python},
   url={http://dx.doi.org/10.1145/3510454.3516829},
   DOI={10.1145/3510454.3516829},
   booktitle={Proceedings of the ACM/IEEE 44th International Conference on Software Engineering: Companion Proceedings},
   publisher={ACM},
   author={Lukasczyk, Stephan and Fraser, Gordon},
   year={2022},
   month=may, pages={168–172},
   collection={ICSE ’22} }

@article{Metamorphic_survey,
author = {Chen, Tsong Yueh and Kuo, Fei-Ching and Liu, Huai and Poon, Pak-Lok and Towey, Dave and Tse, T. H. and Zhou, Zhi Quan},
title = {Metamorphic Testing: A Review of Challenges and Opportunities},
year = {2018},
issue_date = {January 2019},
publisher = {Association for Computing Machinery},
address = {New York, NY, USA},
volume = {51},
number = {1},
issn = {0360-0300},
url = {https://doi.org/10.1145/3143561},
doi = {10.1145/3143561},
journal = {ACM Comput. Surv.},
month = jan,
articleno = {4},
numpages = {27}
}

@inproceedings{benchmark_,
author = {Malviya, Rajesh and Javalkar, Vishal and Malviya, Rano},
year = {2024},
month = {09},
pages = {},
title = {Scalability and Performance Benchmarking of LangChain, LlamaIndex, and Haystack for Enterprise AI Customer Support Systems},
doi = {10.21428/e90189c8.43aeb06e}
}

@misc{aid2024,
      title={LLM-Powered Test Case Generation for Detecting Bugs in Plausible Programs}, 
      author={Kaibo Liu and Zhenpeng Chen and Yiyang Liu and Jie M. Zhang and Mark Harman and Yudong Han and Yun Ma and Yihong Dong and Ge Li and Gang Huang},
      year={2025},
      eprint={2404.10304},
      archivePrefix={arXiv},
      primaryClass={cs.SE},
      url={https://arxiv.org/abs/2404.10304}, 
}

@misc{doc2oracle2024,
      title={Doc2OracLL: Investigating the Impact of Documentation on LLM-based Test Oracle Generation}, 
      author={Soneya Binta Hossain and Raygan Taylor and Matthew Dwyer},
      year={2025},
      eprint={2412.09360},
      archivePrefix={arXiv},
      primaryClass={cs.SE},
      url={https://arxiv.org/abs/2412.09360}, 
}

@misc{candor2024,
      title={Hallucination to Consensus: Multi-Agent LLMs for End-to-End Test Generation with Accurate Oracles}, 
      author={Qinghua Xu and Guancheng Wang and Lionel Briand and Kui Liu},
      year={2025},
      eprint={2506.02943},
      archivePrefix={arXiv},
      primaryClass={cs.SE},
      url={https://arxiv.org/abs/2506.02943}, 
}

@inproceedings{yao2023react,
  title = {{ReAct}: Synergizing Reasoning and Acting in Language Models},
  author = {Yao, Shunyu and Zhao, Jeffrey and Yu, Dian and Du, Nan and Shafran, Izhak and Narasimhan, Karthik and Cao, Yuan},
  booktitle = {International Conference on Learning Representations (ICLR)},
  year = {2023},
  url = {https://arxiv.org/abs/2210.03629},
}

@misc{liu2023lost,
      title={Lost in the Middle: How Language Models Use Long Contexts}, 
      author={Nelson F. Liu and Kevin Lin and John Hewitt and Ashwin Paranjape and Michele Bevilacqua and Fabio Petroni and Percy Liang},
      year={2023},
      eprint={2307.03172},
      archivePrefix={arXiv},
      primaryClass={cs.CL},
      url={https://arxiv.org/abs/2307.03172}, 
}

@misc{luo2025agent,
      title={Large Language Model Agent: A Survey on Methodology, Applications and Challenges}, 
      author={Junyu Luo and Weizhi Zhang and Ye Yuan and Yusheng Zhao and Junwei Yang and Yiyang Gu and Bohan Wu and Binqi Chen and Ziyue Qiao and Qingqing Long and Rongcheng Tu and Xiao Luo and Wei Ju and Zhiping Xiao and Yifan Wang and Meng Xiao and Chenwu Liu and Jingyang Yuan and Shichang Zhang and Yiqiao Jin and Fan Zhang and Xian Wu and Hanqing Zhao and Dacheng Tao and Philip S. Yu and Ming Zhang},
      year={2025},
      eprint={2503.21460},
      archivePrefix={arXiv},
      primaryClass={cs.CL},
      url={https://arxiv.org/abs/2503.21460}, 
}

@misc{xi2023agent,
      title={The Rise and Potential of Large Language Model Based Agents: A Survey}, 
      author={Zhiheng Xi and Wenxiang Chen and Xin Guo and Wei He and Yiwen Ding and Boyang Hong and Ming Zhang and Junzhe Wang and Senjie Jin and Enyu Zhou and Rui Zheng and Xiaoran Fan and Xiao Wang and Limao Xiong and Yuhao Zhou and Weiran Wang and Changhao Jiang and Yicheng Zou and Xiangyang Liu and Zhangyue Yin and Shihan Dou and Rongxiang Weng and Wensen Cheng and Qi Zhang and Wenjuan Qin and Yongyan Zheng and Xipeng Qiu and Xuanjing Huang and Tao Gui},
      year={2023},
      eprint={2309.07864},
      archivePrefix={arXiv},
      primaryClass={cs.AI},
      url={https://arxiv.org/abs/2309.07864}, 
}

@article{Wang2024agent,
   title={A survey on large language model based autonomous agents},
   volume={18},
   ISSN={2095-2236},
   url={http://dx.doi.org/10.1007/s11704-024-40231-1},
   DOI={10.1007/s11704-024-40231-1},
   number={6},
   journal={Frontiers of Computer Science},
   publisher={Springer Science and Business Media LLC},
   author={Wang, Lei and Ma, Chen and Feng, Xueyang and Zhang, Zeyu and Yang, Hao and Zhang, Jingsen and Chen, Zhiyuan and Tang, Jiakai and Chen, Xu and Lin, Yankai and Zhao, Wayne Xin and Wei, Zhewei and Wen, Jirong},
   year={2024},
   month=mar }

@article{Ou2026dlfuzz,
   title={MirrorFuzz: Leveraging LLM and Shared Bugs for Deep Learning Framework APIs Fuzzing},
   volume={52},
   ISSN={2326-3881},
   url={http://dx.doi.org/10.1109/TSE.2025.3619966},
   DOI={10.1109/tse.2025.3619966},
   number={1},
   journal={IEEE Transactions on Software Engineering},
   publisher={Institute of Electrical and Electronics Engineers (IEEE)},
   author={Ou, Shiwen and Li, Yuwei and Yu, Lu and Wei, Chengkun and Wen, Tingke and Chen, Qiangpu and Chen, Yu and Tang, Haizhi and Pan, Zulie},
   year={2026},
   month=jan, pages={360–375} }

@ARTICLE{oracle2014,
  author={Barr, Earl T. and Harman, Mark and McMinn, Phil and Shahbaz, Muzammil and Yoo, Shin},
  journal={IEEE Transactions on Software Engineering}, 
  title={The Oracle Problem in Software Testing: A Survey}, 
  year={2015},
  volume={41},
  number={5},
  pages={507-525},
  doi={10.1109/TSE.2014.2372785}}

@misc{huang2024largelanguagemodelsselfcorrect,
      title={Large Language Models Cannot Self-Correct Reasoning Yet}, 
      author={Jie Huang and Xinyun Chen and Swaroop Mishra and Huaixiu Steven Zheng and Adams Wei Yu and Xinying Song and Denny Zhou},
      year={2024},
      eprint={2310.01798},
      archivePrefix={arXiv},
      primaryClass={cs.CL},
      url={https://arxiv.org/abs/2310.01798}, 
}

@inproceedings{2023judge,
author = {Zheng, Lianmin and Chiang, Wei-Lin and Sheng, Ying and Zhuang, Siyuan and Wu, Zhanghao and Zhuang, Yonghao and Lin, Zi and Li, Zhuohan and Li, Dacheng and Xing, Eric P. and Zhang, Hao and Gonzalez, Joseph E. and Stoica, Ion},
title = {Judging LLM-as-a-judge with MT-bench and Chatbot Arena},
year = {2023},
publisher = {Curran Associates Inc.},
address = {Red Hook, NY, USA},
booktitle = {Proceedings of the 37th International Conference on Neural Information Processing Systems},
articleno = {2020},
numpages = {29},
location = {New Orleans, LA, USA},
series = {NIPS '23}
}

@software{Ramirez_FastAPI,
author = {Ramírez, Sebastián},
license = {MIT},
title = {{FastAPI}},
url = {https://github.com/fastapi/fastapi}
}

@INPROCEEDINGS{codamosa,
  author={Lemieux, Caroline and Inala, Jeevana Priya and Lahiri, Shuvendu K. and Sen, Siddhartha},
  booktitle={2023 IEEE/ACM 45th International Conference on Software Engineering (ICSE)}, 
  title={CodaMosa: Escaping Coverage Plateaus in Test Generation with Pre-trained Large Language Models}, 
  year={2023},
  volume={},
  number={},
  pages={919-931},
  doi={10.1109/ICSE48619.2023.00085}}

@article{telpa,
author = {Yang, Chen and Chen, Junjie and Lin, Bin and Wang, Ziqi and Zhou, Jianyi},
title = {Advancing Code Coverage: Incorporating Program Analysis with Large Language Models},
year = {2025},
publisher = {Association for Computing Machinery},
address = {New York, NY, USA},
issn = {1049-331X},
url = {https://doi.org/10.1145/3748505},
doi = {10.1145/3748505},
note = {Just Accepted},
journal = {ACM Trans. Softw. Eng. Methodol.},
month = jul
}

@misc{agent4se,
      title={Understanding Software Engineering Agents: A Study of Thought-Action-Result Trajectories}, 
      author={Islem Bouzenia and Michael Pradel},
      year={2025},
      eprint={2506.18824},
      archivePrefix={arXiv},
      primaryClass={cs.SE},
      url={https://arxiv.org/abs/2506.18824}, 
}

@inproceedings{swe-agent,
author = {Yang, John and Jimenez, Carlos E. and Wettig, Alexander and Lieret, Kilian and Yao, Shunyu and Narasimhan, Karthik and Press, Ofir},
title = {SWE-agent: agent-computer interfaces enable automated software engineering},
year = {2024},
isbn = {9798331314385},
publisher = {Curran Associates Inc.},
address = {Red Hook, NY, USA},
articleno = {1601},
numpages = {125},
location = {Vancouver, BC, Canada},
series = {NIPS '24}
}

@misc{swe-bench,
      title={SWE-bench: Can Language Models Resolve Real-World GitHub Issues?}, 
      author={Carlos E. Jimenez and John Yang and Alexander Wettig and Shunyu Yao and Kexin Pei and Ofir Press and Karthik Narasimhan},
      year={2024},
      eprint={2310.06770},
      archivePrefix={arXiv},
      primaryClass={cs.CL},
      url={https://arxiv.org/abs/2310.06770}, 
}

@inproceedings{toolformer,
author = {Schick, Timo and Dwivedi-Yu, Jane and Dess\'{\i}, Roberto and Raileanu, Roberta and Lomeli, Maria and Hambro, Eric and Zettlemoyer, Luke and Cancedda, Nicola and Scialom, Thomas},
title = {Toolformer: language models can teach themselves to use tools},
year = {2023},
publisher = {Curran Associates Inc.},
address = {Red Hook, NY, USA},
booktitle = {Proceedings of the 37th International Conference on Neural Information Processing Systems},
articleno = {2997},
numpages = {13},
location = {New Orleans, LA, USA},
series = {NIPS '23}
}

@misc{metagpt,
      title={MetaGPT: Meta Programming for A Multi-Agent Collaborative Framework}, 
      author={Sirui Hong and Mingchen Zhuge and Jiaqi Chen and Xiawu Zheng and Yuheng Cheng and Ceyao Zhang and Jinlin Wang and Zili Wang and Steven Ka Shing Yau and Zijuan Lin and Liyang Zhou and Chenyu Ran and Lingfeng Xiao and Chenglin Wu and Jürgen Schmidhuber},
      year={2024},
      eprint={2308.00352},
      archivePrefix={arXiv},
      primaryClass={cs.AI},
      url={https://arxiv.org/abs/2308.00352}, 
}

@misc{toolllm,
      title={ToolLLM: Facilitating Large Language Models to Master 16000+ Real-world APIs}, 
      author={Yujia Qin and Shihao Liang and Yining Ye and Kunlun Zhu and Lan Yan and Yaxi Lu and Yankai Lin and Xin Cong and Xiangru Tang and Bill Qian and Sihan Zhao and Lauren Hong and Runchu Tian and Ruobing Xie and Jie Zhou and Mark Gerstein and Dahai Li and Zhiyuan Liu and Maosong Sun},
      year={2023},
      eprint={2307.16789},
      archivePrefix={arXiv},
      primaryClass={cs.AI},
      url={https://arxiv.org/abs/2307.16789}, 
}

@inproceedings{intercode,
author = {Yang, John and Prabhakar, Akshara and Narasimhan, Karthik and Yao, Shunyu},
title = {InterCode: standardizing and benchmarking interactive coding with execution feedback},
year = {2023},
publisher = {Curran Associates Inc.},
address = {Red Hook, NY, USA},

booktitle = {Proceedings of the 37th International Conference on Neural Information Processing Systems},
articleno = {1035},
numpages = {29},
location = {New Orleans, LA, USA},
series = {NIPS '23}
}

@INPROCEEDINGS{misuse,
  author={Li, Xia and Jiang, Jiajun and Benton, Samuel and Xiong, Yingfei and Zhang, Lingming},
  booktitle={2021 14th IEEE Conference on Software Testing, Verification and Validation (ICST)}, 
  title={A Large-scale Study on API Misuses in the Wild}, 
  year={2021},
  volume={},
  number={},
  pages={241-252},
  doi={10.1109/ICST49551.2021.00034}}

@misc{stateflow,
      title={StateFlow: Enhancing LLM Task-Solving through State-Driven Workflows}, 
      author={Yiran Wu and Tianwei Yue and Shaokun Zhang and Chi Wang and Qingyun Wu},
      year={2024},
      eprint={2403.11322},
      archivePrefix={arXiv},
      primaryClass={cs.CL},
      url={https://arxiv.org/abs/2403.11322}, 
}

@misc{chattester,
      title={No More Manual Tests? Evaluating and Improving ChatGPT for Unit Test Generation}, 
      author={Zhiqiang Yuan and Yiling Lou and Mingwei Liu and Shiji Ding and Kaixin Wang and Yixuan Chen and Xin Peng},
      year={2024},
      eprint={2305.04207},
      archivePrefix={arXiv},
      primaryClass={cs.SE},
      url={https://arxiv.org/abs/2305.04207}, 
}

@article{liu-etal-2024-lost,
    title = "Lost in the Middle: How Language Models Use Long Contexts",
    author = "Liu, Nelson F.  and
      Lin, Kevin  and
      Hewitt, John  and
      Paranjape, Ashwin  and
      Bevilacqua, Michele  and
      Petroni, Fabio  and
      Liang, Percy",
    journal = "Transactions of the Association for Computational Linguistics",
    volume = "12",
    year = "2024",
    address = "Cambridge, MA",
    publisher = "MIT Press",
    url = "https://aclanthology.org/2024.tacl-1.9/",
    doi = "10.1162/tacl_a_00638",
    pages = "157--173",
   
}

@inproceedings{tricorder,
author = {Sadowski, Caitlin and van Gogh, Jeffrey and Jaspan, Ciera and S\"{o}derberg, Emma and Winter, Collin},
title = {Tricorder: building a program analysis ecosystem},
year = {2015},
isbn = {9781479919345},
publisher = {IEEE Press},
booktitle = {Proceedings of the 37th International Conference on Software Engineering - Volume 1},
pages = {598–608},
numpages = {11},
location = {Florence, Italy},
series = {ICSE '15}
}

@misc{nexas,
      title={Nexus: Execution-Grounded Multi-Agent Test Oracle Synthesis}, 
      author={Dong Huang and Mingzhe Du and Jie M. Zhang and Zheng Lin and Meng Luo and Qianru Zhang and See-Kiong Ng},
      year={2025},
      eprint={2510.26423},
      archivePrefix={arXiv},
      primaryClass={cs.SE},
      url={https://arxiv.org/abs/2510.26423}, 
}

@inproceedings{xue_agentstudy,
author = {Xue, Ziluo and Zhao, Yanjie and Wang, Shenao and Chen, Kai and Wang, Haoyu},
title = {A Characterization Study of Bugs in LLM Agent Workflow Orchestration Frameworks},
year = {2025},
publisher = {IEEE Press},
url = {https://doi.org/10.1109/ASE63991.2025.00278},
doi = {10.1109/ASE63991.2025.00278},
booktitle = {2025 40th IEEE/ACM International Conference on Automated Software Engineering (ASE)},
pages = {3369–3380},
numpages = {12},
location = {Seoul, Korea, Republic of}
}

@inproceedings{zhang2025icml-agent,
  title     = {{Which Agent Causes Task Failures and When? on Automated Failure Attribution of LLM Multi-Agent Systems}},
  author    = {Zhang, Shaokun and Yin, Ming and Zhang, Jieyu and Liu, Jiale and Han, Zhiguang and Zhang, Jingyang and Li, Beibin and Wang, Chi and Wang, Huazheng and Chen, Yiran and Wu, Qingyun},
  booktitle = {Proceedings of the 42nd International Conference on Machine Learning},
  year      = {2025},
  pages     = {76583-76599},
  volume    = {267},
  url       = {https://mlanthology.org/icml/2025/zhang2025icml-agent/}
}

\end{document}